\documentclass[nofootinbib,preprint,superscriptaddress, 11pt]{revtex4}
\pdfoutput=1
\usepackage[usenames,dvipsnames]{color}
\usepackage{graphicx}
\usepackage{amsmath}
\usepackage{amsfonts}
\usepackage{amssymb}
\usepackage{dsfont}
\usepackage{dcolumn}
\usepackage{bm}
\usepackage{slashed}
\usepackage{pstricks}
\usepackage{ulem}

\newcolumntype{x}[1]{
>{\centering}p{#1}}%

\newcommand{\GeV}      {~\mathrm{GeV}}

\def \cha{\widetilde{\chi}^{\pm}_1}

\newcommand{\beqn}{\begin{eqnarray}}
\newcommand{\eeqn}{\end{eqnarray}}
\newcommand{\be}{\begin{equation}}
\newcommand{\ee}{\end{equation}}

\newcommand{\mathsym}[1]{{}}

\def \cha{\tilde{\chi}^{\pm}_1}

\def \na{\tilde{\chi}^{0}_1}

\def \n34{\tilde{\chi}^{0}_{3,4}}

\def \ta{\tilde{t}_1}

\def \sta{\tilde{\tau}_1}

\def\met100{\slashed{E}_T\geq 100 \GeV}

\def\beq{\begin{eqnarray}}
\def\eeq{\end{eqnarray}}
\def\CR{\nonumber \\ }

\begin{document}

\baselineskip=21pt
\begin{center}
UTTG-16-11
\hspace{0.3cm} TCC-018-11
\hspace{0.3cm} MCTP-11-27
\end{center}

\title{Probing EWSB Naturalness in Unified SUSY Models with Dark Matter}

\author{Stephen Amsel}
\email{samsel@umich.edu}
\affiliation{Michigan Center for Theoretical Physics, Department of Physics,
University of Michigan, Ann Arbor, MI 48109, USA}
\author{Katherine Freese}
\email{ktfreese@umich.edu}
\affiliation{Michigan Center for Theoretical Physics, Department of Physics, 
University of Michigan, Ann Arbor, MI 48109, USA}
\affiliation{Texas Cosmology Center, University of Texas, Austin, TX 78712, USA}
\author{Pearl Sandick}
\email{pearl@physics.utexas.edu}
\affiliation{Texas Cosmology Center, University of Texas, Austin, TX 78712, USA}
\affiliation{Theory Group, Department of Physics, University of Texas, Austin, TX 78712, USA}
\affiliation{Department of Physics and Astronomy, University of Utah, Salt Lake City, UT 84112, USA}


\begin{abstract}
We have studied Electroweak Symmetry Breaking (EWSB) fine-tuning in the context of two unified Supersymmetry scenarios:  the Constrained Minimal Supersymmetric Model (CMSSM) and models with Non-Universal Higgs Masses (NUHM), in light of current and upcoming direct
detection dark matter experiments.  We consider both those models that satisfy a one-sided bound on the relic density of neutralinos, $\Omega_{\na}h^2 < 0.12$, and also the subset that satisfy the two-sided bound in which the relic density is within the 2 sigma best fit of WMAP7 + BAO + H0 data.  
We find that current direct searches for dark matter probe the least fine-tuned regions of parameter-space, or equivalently those of lowest $\mu$, and will tend to probe progressively more and more fine-tuned models, though the trend is more pronounced in the CMSSM than in the NUHM.
Additionally, we examine several subsets of model points, categorized by common mass hierarchies; $M_{\na} \sim M_{\cha}, M_{\na} \sim M_{\sta}, M_{\na} \sim M_{\ta}$, the light and heavy Higgs poles, and any additional models classified as ``other''; the relevance of these mass
hierarchies is their connection to the preferred neutralino annihilation channel that determines the 
relic abundance.  For each of these subsets of models we investigated the degree of fine-tuning and discoverability in current and next generation direct detection experiments.
\end{abstract}

\maketitle

\section{Introduction}

The Minimal Supersymmetric Standard Model (MSSM) is the simplest
supersymmetric extension of particle physics beyond the Standard Model.
If supersymmetry is broken near the weak scale, not only
is the MSSM a framework in which 
gauge coupling unification can be achieved~\cite{gut}, but it also provides a compelling candidate for particle dark matter~\cite{cdm}; the lightest supersymmetric particle (LSP), which is expected to be stable in many supersymmetric realizations.
One of the most simple and oft-studied MSSM realizations is the constrained MSSM (CMSSM)~\cite{cmssm}, in which the entire spectrum of particles and their interactions are specified at some high input scale, typically the supersymmetric GUT scale, by four free parameters and a sign: a universal mass for all gauginos, $M_{1/2}$; a universal mass for all scalars, $M_0$; a universal value for the trilinear couplings, $A_0$; the ratio of the Higgs vacuum expectation values, $\tan\beta$; and the sign of the Higgs mixing parameter, $\mu$.  However, it is by no means necessary that all scalar masses are unified at a high scale.  In fact, the soft supersymmetry-breaking contributions to the Higgs scalar masses
are generally not related to the squark and slepton masses, even in the context of SUSY GUTs\footnote{We note that in the minimal supergravity (mSUGRA) framework this full universality of scalar masses does occur, but it is absent in more general effective supergravity theories~\cite{Cremmer:1983bf, Ellis:1983sf}.}.

In this paper we investigate the relationship between electroweak naturalness, direct dark matter detection prospects, and the mass hierarchy of supersymmetric particles in two unified variants of the MSSM: a case with full universality of scalar masses at the GUT scale, 
the CMSSM, 
and a case in which the supersymmetry-breaking contributions to the scalar masses of the MSSM Higgs multiplets are allowed to deviate from the universal value of the squark and slepton masses at the GUT scale, models with Non-Universal Higgs Masses (NUHM)~\cite{nuhm}. While it is possible that the supersymmetry-breaking contributions to the scalar Higgs masses themselves are universal at the GUT scale (often called NUHM1, for the one additional free parameter required to specify the model)~\cite{nuhm1}, here we examine the more general case that the two Higgs masses are unrelated (commonly referred to as NUHM2), of which the NUHM1 is a subset.
In both the CMSSM and the NUHM, the dark matter candidate is
the lightest neutralino, which is a linear combination of the supersymmetric
partners of the photon, the Z boson, and the neutral scalar Higgs particles.  Neutralino LSPs are excellent dark matter candidates, possessing roughly the right annihilation cross section and mass to account for the
observed density of cold dark matter in the universe, assuming they are thermal relics.  
According
to the analysis in \cite{WMAP7}, the cold dark matter density has the value
\begin{equation}
\label{eq:relic}
\Omega_{CDM} h^2 = 0.1127 \pm 0.0036,
\end{equation}
where 
$h$ is the Hubble constant $H_0$ in units of 100 km/s/Mpc,
$\Omega_{CDM} = \rho_{CDM}/\rho_c$ is the fraction of the
dark matter density in units of the critical density
$\rho_c = 3H_0^2/(8\pi G) \sim 10^{-29}$ g/cm$^3$,
and the best fit and $1\sigma$ errors are obtained from
a combination of WMAP7, BAO, and H0 data.

Despite the successes of the MSSM, fine-tuning of the Z mass is a generic issue for supersymmetric models.
Neglecting loop corrections, the Z mass in the MSSM is given by 
\beq 
\label{zmass}
m_Z^2 =
\frac{|m_{H_d}^2-m_{H_u}^2|}{\sqrt{1-\sin^2 2\beta}}-m_{H_d}^2-m_{H_u}^2-2|\mu|^2,
\eeq
where $m_{H_u}$ and $m_{H_d}$ are the SUSY-breaking contributions to the effective masses of the up- and down-type Higgs fields, respectively, and all parameters are defined at $m_Z$.
Clearly, a cancellation of the terms on the right hand side
is required in order to obtain the measured value of $m_Z$, a particularly unnerving situation given that typical values for parameters on the right hand side can be orders of magnitude from the weak scale. 

As noted in~\cite{Ellis:1986yg} and~\cite{BarbieriGiudice}, the degree of fine-tuning may be quantified using log-derivatives.  Here, we follow~\cite{Perelstein:2007nx} and compute the quantity 
\beq 
A(\xi)\,=\,\left| \frac{\partial\log m_Z^2}{\partial\log \xi}\right|,
\label{eq:ftpars} 
\eeq
where $\xi=m_{H_u}^2$, $m_{H_d}^2$, $b$, and $\mu$ are the relevant Lagrangian parameters.  
Then 
\beqn
 A(\mu) &=&
\frac{4\mu^2}{m_Z^2}\,\left(1+\frac{m_A^2+m_Z^2}{m_A^2} \tan^2 2\beta
\right), \CR A(b) &=& \left( 1+\frac{m_A^2}{m_Z^2}\right)\tan^2
2\beta, \CR A(m_{H_u}^2) &=& \left| \frac{1}{2}\cos2\beta
+\frac{m_A^2}{m_Z^2}\cos^2\beta
-\frac{\mu^2}{m_Z^2}\right|\times\left(1-\frac{1}{\cos2\beta}+
\frac{m_A^2+m_Z^2}{m_A^2} \tan^2 2\beta \right), \CR A(m_{H_d}^2) &=&
\left| -\frac{1}{2}\cos2\beta +\frac{m_A^2}{m_Z^2}\sin^2\beta
-\frac{\mu^2}{m_Z^2}\right|\times\left|1+\frac{1}{\cos2\beta}+
\frac{m_A^2+m_Z^2}{m_A^2} \tan^2 2\beta \right|, \CR 
\eeqn
where it is assumed that $\tan\beta>1$. 
The overall fine-tuning $\Delta$ is
defined as
\beq
\Delta = \sqrt{A(\mu)^2+A(b)^2+A(m_{H_u}^2)^2+A(m_{H_d}^2)^2},
\label{eq:delta}
\eeq
with values of $\Delta$ far above one indicating significant fine-tuning. Quantum corrections further contribute to the fine-tuning, 
e.g.~the one-loop contribution to the $m_{H_u}^2$ parameter from top and stop loops.

\begin{figure}[h]
\includegraphics[width=80mm, height=80mm]{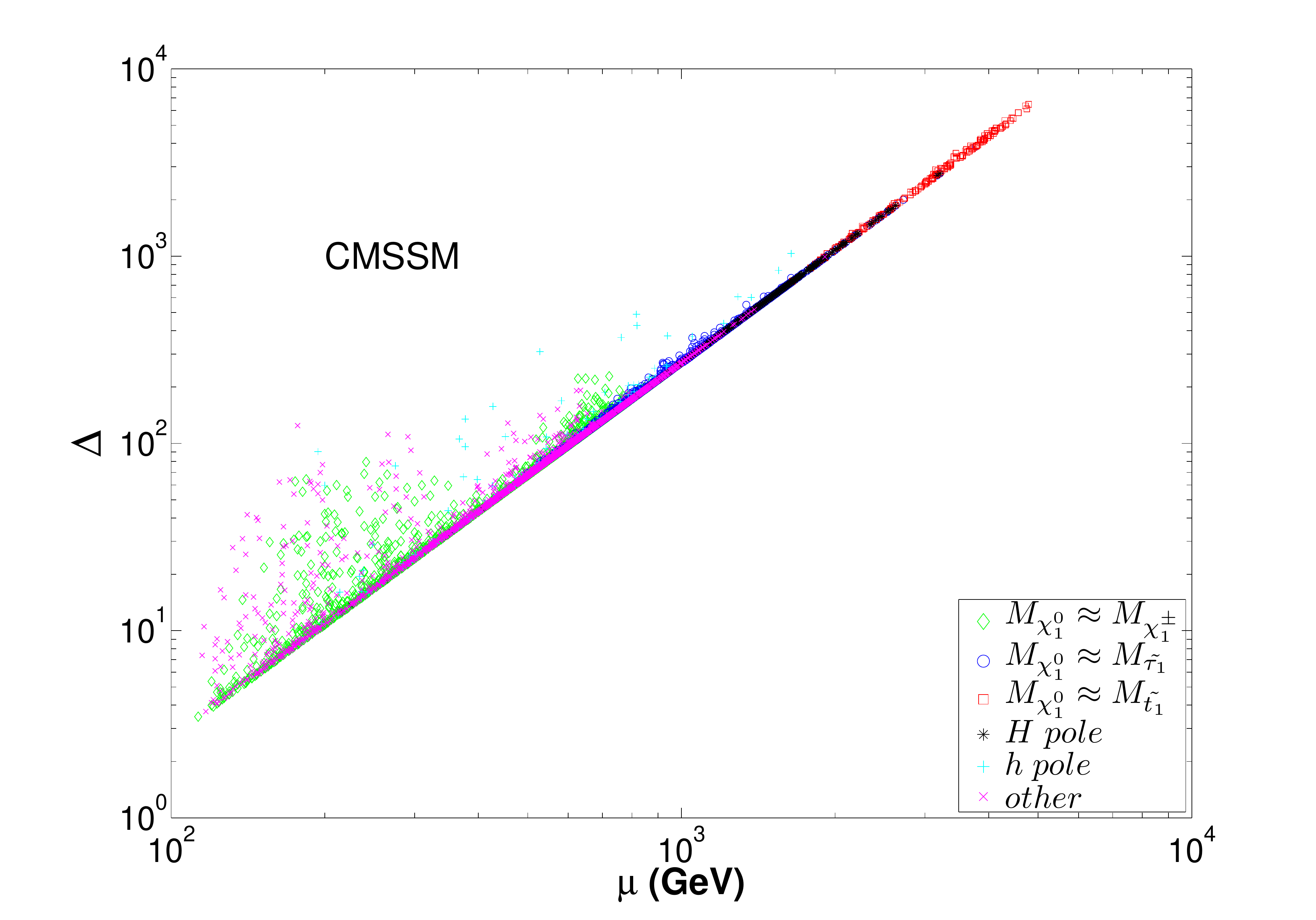}
\includegraphics[width=80mm, height=80mm]{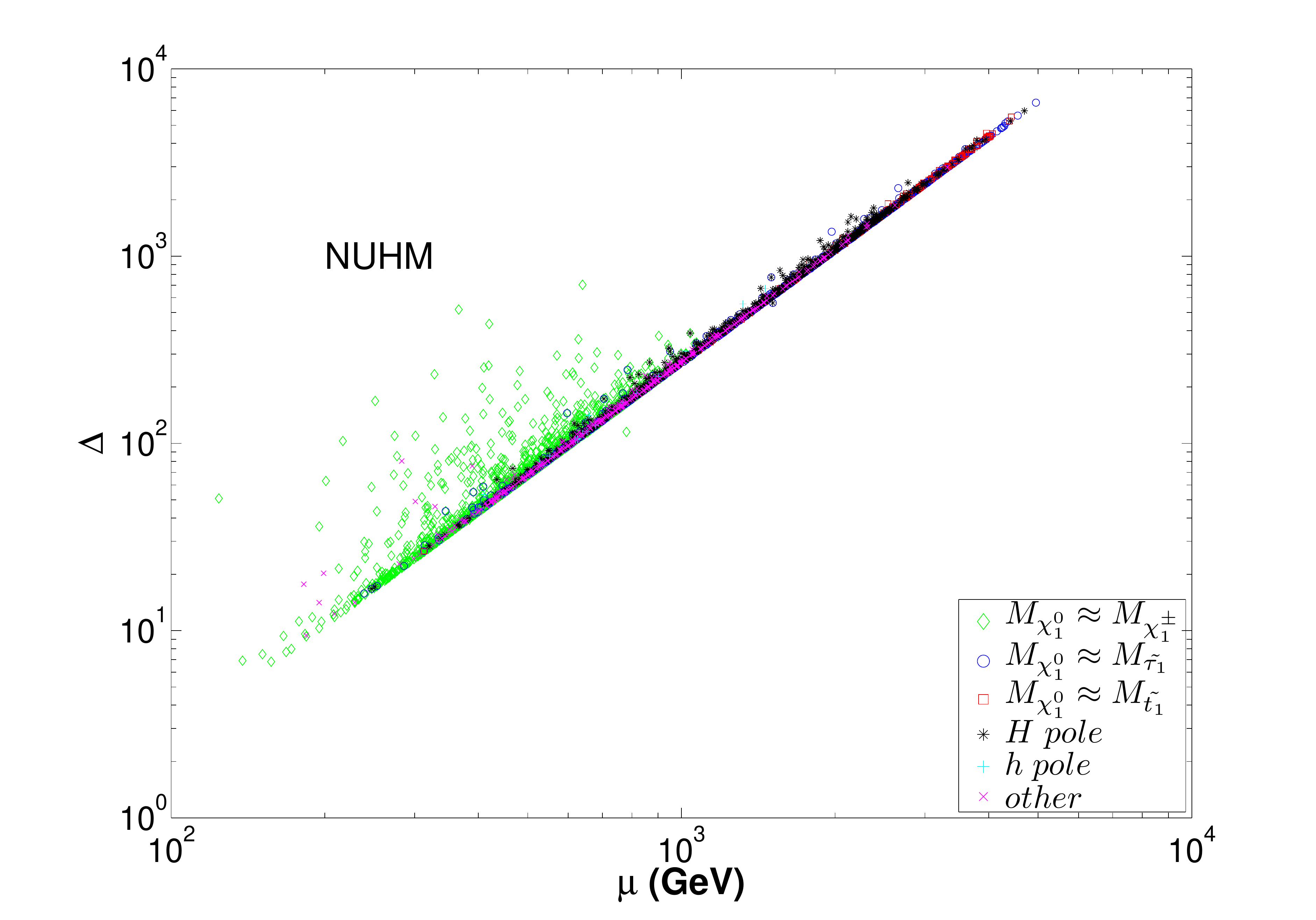}
\caption{Fine-tuning, parametrized by $\Delta$, plotted as a function of $\mu$ in both the CMSSM and NUHM for
$0<{\rm tan}\beta <60$. Models are color-coded by their mass
hierarchy as indicated in the legend.
A one-sided bound $\Omega_{\na}h^2 < 0.12$ has been applied.}
\label{delta_mu}
\end{figure}

In this paper we use the MicrOMEGAs code~\cite{micromegas} with SUSPECT~\cite{suspect} to compute the fine-tuning
parameter $\Delta$ (accurate to at least one-loop). We note that bounds on $m_{\cha}$ imply $\mu > 100$ GeV.  As illustrated in Fig.~\ref{delta_mu}, $\Delta$ is strongly correlated with $\mu$.  
This relationship between $\Delta$ and $\mu$ can be easily understood by considering the approximation
\beq
\Delta = \sqrt{5} \times \frac{\mu^{2}}{m^{2}_{Z}} + O(\frac{1}{\tan^{2}\beta}) ,
\label{eq_deltamu}
\eeq
valid at large $\tan\beta$.  Throughout this paper, however, we use the full calculation of Eq.~\ref{eq:delta}.

The point of this paper is to 
study the amount of  fine-tuning in the CMSSM and
NUHM under the assumption that the lightest neutralino makes up some portion of the dark matter in the Universe, with a focus on the relationship between fine-tuning and prospects for direct detection of dark matter in these scenarios.
Direct searches for dark matter seek to detect the scattering of dark matter particles off of 
nuclei in low-background detectors.  Many such searches are being pursued, among them~\cite{Aprile:2011hi, Ahmed:2010wy, Ahmed:2011gh, Hall:2010zz, Behnke:2010xt, Bernabei:2010mq, Behnke:2008zz, Aalseth:2010vx, Angloher:2002in, Stodolsky:2004dsu, Lin:2007ka}.  For brevity, here we consider the 
current bounds and future prospects specifically for the XENON experiment only.  Current bounds
have been presented for 100 live days of operation of the XENON-100 detector~\cite{Aprile:2011hi}, while future projections are for the ton-scale detector, XENON-1T~\cite{Aprile:2009yh}.  Specifically, we apply
the latest bounds from the XENON-100 experiment on the spin-independent cross section, $\sigma_{SI}$, normalized to scattering off protons (i.e.~we divide out the dependence on the atomic number of the nucleus with which the scattering takes place).  We note that although the discussion is focused on the XENON detectors, the cross sections we present are not specific to any particular experiment.  

Fine-tuning has long been a concern for phenomenological models within the MSSM framework. The sensitivity of the neutralino dark matter abundance to fine-tuning of the CMSSM inputs was studied in~\cite{Ellis:2001zk}, while EWSB and dark matter fine-tuning in the MSSM with non-universal gaugino and third generation scalar masses was studied in~\cite{King:2006tf}.  
The connection between electroweak naturalness and neutralino-nucleus elastic scattering was explored in~\cite{Kitano:2006ws}.
Most recently,~\cite{Cassel:FineTuning} examined the LHC signatures and direct dark matter search prospects for CMSSM models with low fine-tuning, and~\cite{Farina:2011bh} studied fine-tuning in light of recent XENON-100 and LHC constraints.  
As we were completing this manuscript, we became aware also of~\cite{Perelstein:2011tg}, in which the relationship between electroweak fine-tuning and the neutralino-nucleus elastic scattering cross section is also discussed in the context of the MSSM with relevant parameters specified at the weak scale and with the assumption that neutralinos constitute all of the dark matter in the Universe.  Our results are in agreement with their findings.  In this paper, we also study the mass hierarchy of relevant supersymmetric particles as described below.

We assume a thermal history for the LSP and require its  relic density 
 to be less than or equal to that of the cosmological dark matter.  For a predominantly bino-like LSP,  generic annihilation channels do  not in general reduce the relic density sufficiently to meet constraints set by observations of the dark matter density. Co-annihilation with another particle ($\cha$, $\ta$, or $\sta$) or enhancement of the annihilation cross-section by a light or heavy Higgs pole is often necessary for such LSPs. 
We study each of these channels separately by categorizing 
models based on the mass hierarchy of SUSY particles in each: we label them according to the near-degeneracy of the neutralino LSP with the next-to-lightest SUSY particle (NLSP), or by the near-resonance that enhances the LSP annihilation rate.  The categories we consider are near-degeneracy of the LSP with  $\cha$, $\ta$, or $\sta$ 
particles, and $h$- and $A$-pole resonances.  We note that if the LSP has a significant higgsino admixture, it is possible for the relic density of neutralinos to be cosmologically viable even in the absence of a resonance or co-annihilations, and we make no a priori assumptions about the composition of the neutralino LSP.  Mass hierarchies have been studied with respect to spin independent neutralino-nucleon elastic scattering in~\cite{Feldman:MassHierarchies}.  Here we present a simplified categorization scheme in order to focus on the fine-tuning and implications for direct dark matter searches.

For all scans, we take the top mass to be $m_{t} = 173.1$ GeV~\cite{TopMass}.   In both the CMSSM and the NUHM, we assume $\mu >0$ and scan the ranges $1 < \tan\beta < 60$ and $-12$ TeV$ < A_{0} < 12$ TeV. In the CMSSM, we scan $0 < M_{0} < 4$ TeV and $0 < M_{1/2} < 2$ TeV while in NUHM space we scan $0 < M_{0} < 3$ TeV, $0 < M_{1/2} < 2$ TeV, and the GUT-scale Higgs scalar mass parameters $-3$ TeV $< M_{H_{u,d}}(M_{GUT}) < 3$ TeV. We note that in the NUHM, the scan was divided into a more dense scan for $0 < M_{1/2} < 1$ TeV, and a less dense scan for $1$ TeV$ < M_{1/2} < 2$ TeV. The motivation for this division is that lower $M_{1/2}$ implies lower gaugino masses and therefore potentially interesting LHC phenomenology to be explored in follow-up work.  The non-uniform scan does not affect the conclusions of this study, and we would like to note that the sparseness of points should not be taken as an indication of the sparseness of the parameter space.
The assumption of gaugino universality is not relaxed here. Hence, the running of the gaugino masses (calculated using the Renormalization Group Equations of the MSSM) results in the standard rough relations of 1:2:6 for $M_{1} : M_{2} : M_{3}$ at the electroweak scale in both the CMSSM and the NUHM.

\section{Constraints}
\label{Constraints}
\subsection{Accelerator Constraints}
We impose a lower limit on the mass of the light CP-even Higgs boson, $m_{h} > 114$ GeV~\cite{LEPhiggs}.  All accelerator bounds on SUSY parameters were enforced, including $m_{\cha} > 104$ GeV~\cite{Abbiendi:2003sc} and, following~\cite{Feldman:2009zc}, $m_{\ta, \sta} > 100$ GeV.  As in~\cite{Feldman:2009zc}, we take the recommendation of the HFAG~\cite{hfag} (including results from BABAR~\cite{babar}, Belle~\cite{belle}, and CLEO~\cite{cleo}) as well as the updated Standard Model calculation~\cite{bsgSM}, and allow the $3\sigma$ range $2.77\times 10^{-4}<{\mathcal Br}(b\rightarrow s\gamma) <4.27\times 10^{-4}$.  From CDF bounds we require ${\mathcal Br}( B_s \to \mu^{+}\mu^{-})  < 10^{-7}$~\cite{CDFbmumu}.  Finally, we follow Djouadi, Drees, and Kneur to demand $-11.4\times 10^{-10}< \delta(g_{\mu}-2)<9.4\times 10^{-9}$~\cite{Djouadi:2006be}.
 
\subsection{Relic Density}
A thermal cosmological history is assumed. Throughout this study, we apply an upper limit of 
 $\Omega_{\na} h^{2} < 0.12$ for all models in each scan.
 In the penultimate section, however, 
 we further restrict our inquiry to those models 
 with neutralinos providing the correct relic density in the range $0.105 < \Omega_{\na} h^{2} < 0.12$ 
 from Eq. (\ref{eq:relic}) to two sigma.

In the following analysis, we differentiate among SUSY mass hierarchies.
For those cases in which the lightest neutralino is nearly
degenerate in mass with another SUSY particle, we label the models according to the near-degeneracy: 
\beq
m_{\ta}  - m_{\na} < 0.2\, m_{\na},\\  
m_{\cha}  - m_{\na} < 0.15\,  m_{\na},\\
m_{\sta}  - m_{\na} < 0.2\,  m_{\na}.
\eeq
Often (but not always) this corresponds
 to the case of coannihilation of the LSP with the near-degenerate particle as the primary mechanism for  producing the correct relic abundance. 
Cases with 
\beq
\frac{m_{A}}{2} - m_{\na} < 0.1\, m_{\na}\,\, \textrm{ or }\,\,
\frac{m_{h}}{2} - m_{\na} < 0.1\, m_{\na}
\eeq
are labeled as heavy Higgs pole or light Higgs pole 
respectively.  The neutralino LSP associated with the light Higgs pole must have $m_{\na} \sim 50-60$ GeV to 
be compatible with the current limit on the Higgs mass.
Again, most (but not all) of the models in this category have annihilation via a Higgs pole resonance as 
the primary mechanism for producing a small enough relic abundance.  A small subset of the models presented here satisfy a near-degeneracy criterion and the heavy Higgs pole criterion; these models are labeled as having both 
mechanisms\footnote{No models can satisfy both the light Higgs pole criterion as well as near mass degeneracy, 
since $m_{\ta}, m_{\cha} \,\, {\rm or} \,\, m_{\sta}$ as low as 60 GeV is ruled out experimentally.}.
Models not satisfying any of the above criteria as labeled as ``other''.  These include models
where the neutralino LSP has a relatively large Higgsino component,
 which would reduce the relic density regardless of the mass hierarchy.

\section{$(M_{1/2},M_0)$ Plane}

\begin{figure}[h]
\includegraphics[width=81mm, height=70mm]{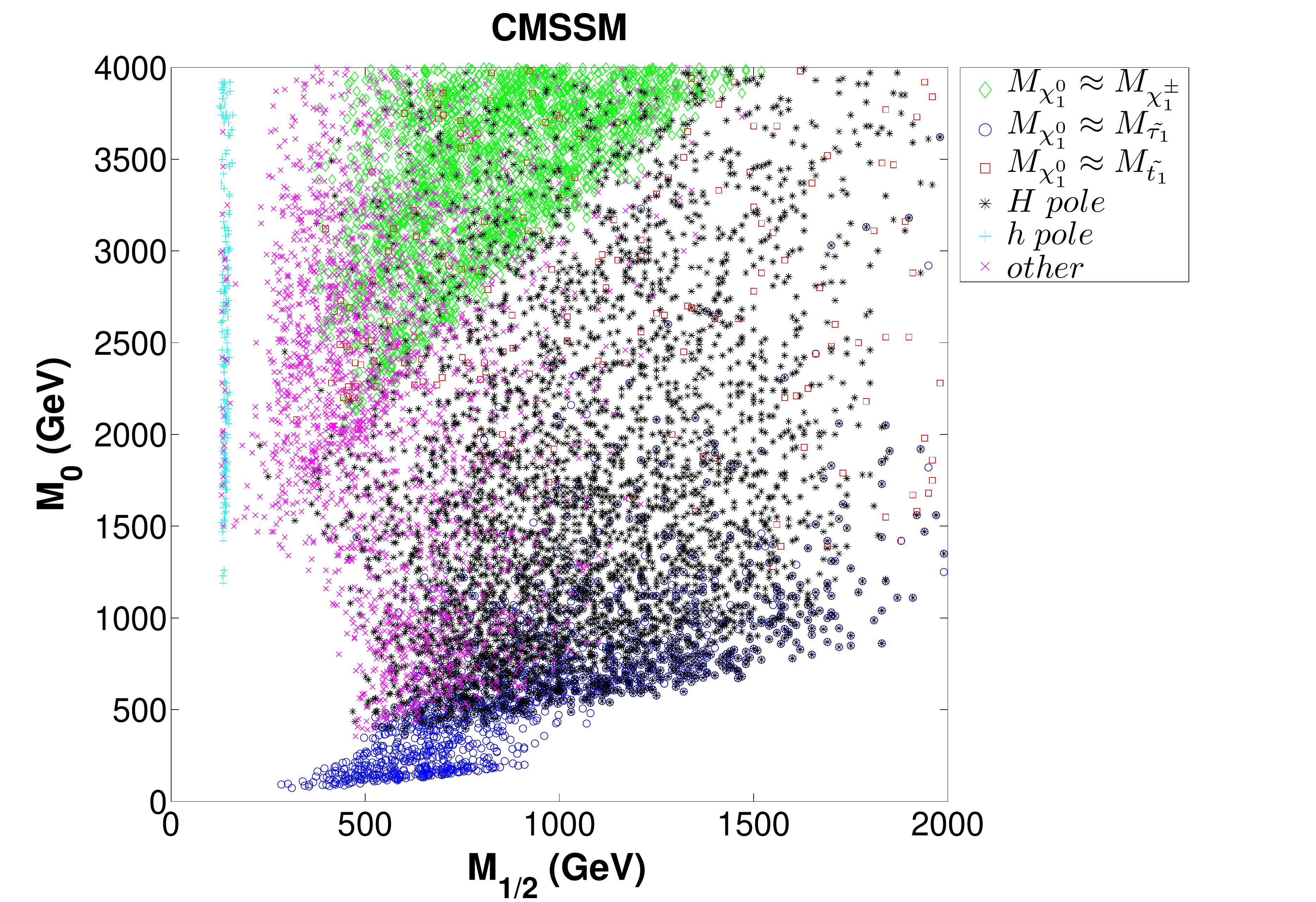}
\includegraphics[width=81mm, height=70mm]{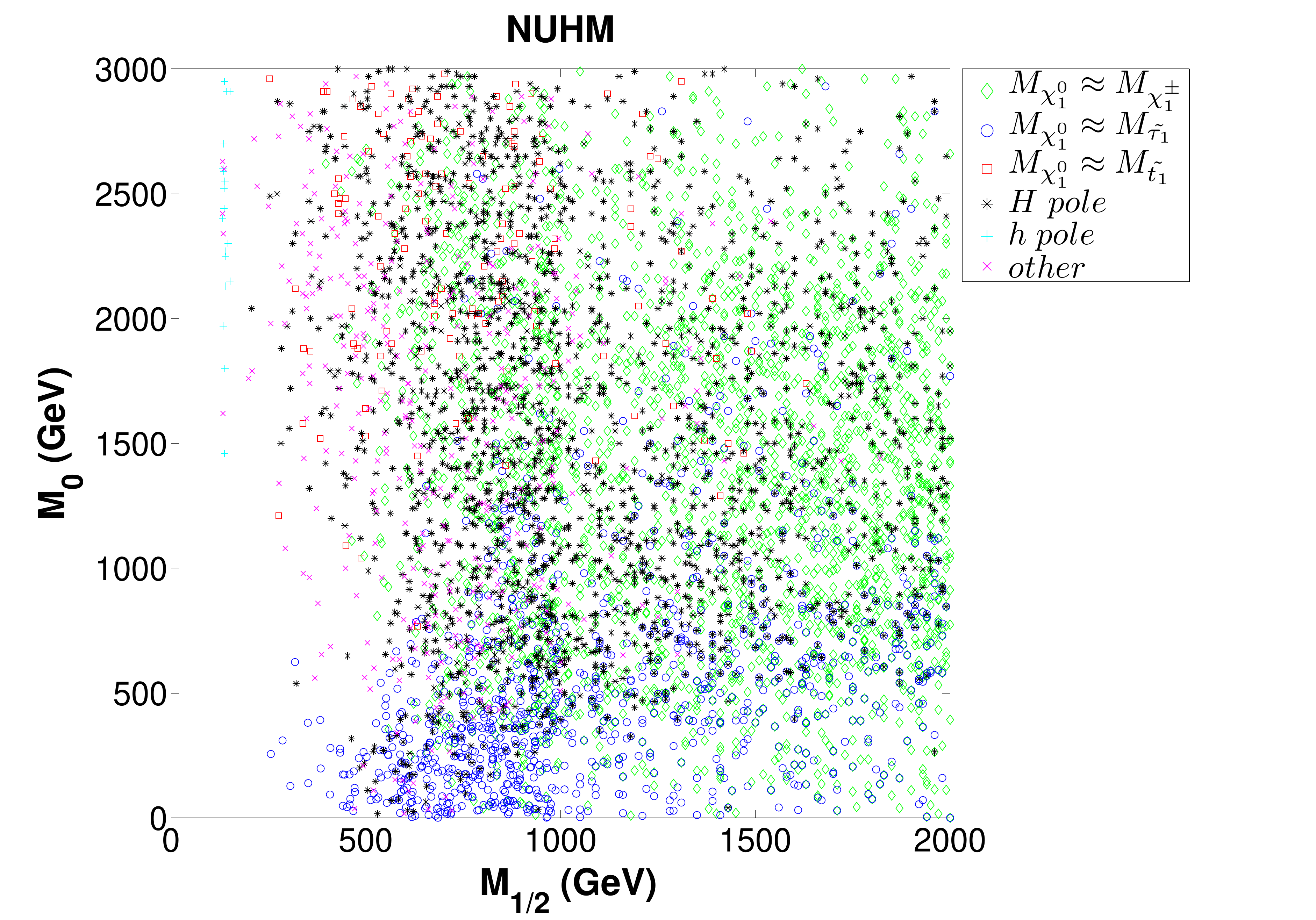}
\caption{The $(M_{1/2},M_0)$ plane of the CMSSM (left) and the NUHM (right).  
Models are color-coded as described in the legend.
Note that the difference in density of the NUHM scan above and below $M_{1/2}=1$ TeV is apparent.  We stress that this is purely an artifact of a scanning choice as described in the text.
A one-sided bound $\Omega_{\na}h^2 < 0.12$ has been applied.} \label{m0_mhf}
\end{figure}

Figure \ref{m0_mhf} illustrates the generalization of the $(M_{1/2},M_0)$ plane from the CMSSM (left)
to the NUHM (right).  
While this has been previously studied 
in the literature, what is new here is the breakdown of the models by mass hierarchy as discussed above:
namely, the models where the lightest neutralino is nearly degenerate with another SUSY particle, the light and heavy
Higgs poles, and ``other''.   
Of the mass hierarchies plotted, some appear more localized in the CMSSM plane than in the NUHM plane.  For example, the $m_{\na}\approx m_{\cha}$ points in the CMSSM all occur at large $M_0$ and small $M_{1/2}$, because that is the only region of the CMSSM plane where the neutralino LSP is significantly higgsino-like so that this near-degeneracy is possible. In the NUHM, however, the GUT-scale restriction that $m_{H_u}=m_{H_d}=M_0$ is relaxed, resulting in significant freedom in the Higgs sector.  As a result, the neutralino LSP may be higgsino-like in any region of the $(M_{1/2},M_0)$ plane.  Indeed, there are $m_{\na}\approx m_{\cha}$ points spread throughout the NUHM plane in the right panel of Fig.~\ref{m0_mhf}.
We remind the reader that
the difference in density of models visible in the right panel of Fig.~\ref{m0_mhf} is due to a difference in density of scans.

\section{Implications of Direct Dark Matter Searches}

In this section we discuss current limits on and projected sensitivity to the CMSSM and NUHM scenarios from the XENON-100 and XENON-1T experiments.  

\subsection{Formalism}
\label{sec:formalism}

The only velocity-independent term in the four-fermion interaction Lagrangian contributing to spin independent scattering of neutralinos with nuclei is
$\mathcal{L}=\alpha_q \bar{\chi} \chi \bar{q} q$~\cite{Falk:1998xj},
with the coefficients $\alpha_q$ calculable from the particle spectrum of the model.
In the zero-momentum-transfer limit, the spin independent elastic scattering cross section for $\na$ scattering on a nucleus with atomic number $Z$ and atomic mass $A$ can be written as
\begin{equation}
\sigma_{SI} = \frac{4 m_r^2}{\pi}\left( Z f_p + (A-Z) f_n\right)^2,
\end{equation}
where $m_r$ is the reduced $\na$-nuclear mass, and the parameters $f_N$ for $N=p$ or $n$ are given by
\begin{equation}
\frac{f_N}{m_N}=\sum_{q=u,d,s} f_{q}^{(N)} \,\frac{\alpha_{q}}{m_q} + \frac{2}{27}\, f_{G}^{(N)}\sum_{q=c,b,t} \frac{\alpha_{q}}{m_q}.
\end{equation}
The nuclear form factors for the light quarks, $f_q^{(N)}$, and the heavy quarks, $f_G^{(N)}$ (induced by gluon exchange), are~\cite{formfactors}
\begin{equation}
m_N f_{q}^{(N)} = \langle N | m_q \bar{q}q | N \rangle = m_q B_q^{(N)},
\end{equation}
and
\begin{equation}
f_{G}^{(N)} = 1- \sum_{q=u,d,s} f_{q}^{(N)}.
\end{equation}
It is useful to parametrize the scattering cross section in terms of the pion-nucleon sigma term, $\sigma_{\pi N}$, and the quantity $\sigma_0$, which are related to the quark masses and $B_q^{(N)}$ by
\begin{equation}
\sigma_{\pi N} = \frac{m_u+m_d}{2}\left(B_u+B_d\right)\,\,\,\, {\rm and} \,\,\,\,
\sigma_0=
\frac{m_u+m_d}{2}\left(B_u+B_d-2B_s\right),
\end{equation}
where we have dropped the superscript $(N)$ due to the relations
\begin{equation}
B_u^{(n)} = B_d^{(p)},\,\,\, B_d^{(n)} = B_u^{(p)},\,\,\, {\rm and}\,\,\, B_s^{(n)} = B_s^{(p)}.
\end{equation}
Finally, introducing the quantities~\cite{zy} 
\begin{equation}
z = \frac{B_u^{(p)}-B_s^{(p)}}{B_d^{(p)}-B_s^{(p)}} = 1.49\,\,\,\, {\rm and}\,\,\, \,
y = \frac{2B_s}{B_u+B_d} = 1-\frac{\sigma_0}{\sigma_{\pi N}},
\end{equation}
the form factors can be written simply as
\begin{eqnarray}
f_{u}^{(N)} &=& \frac{m_uB_u^{(N)}}{m_N} = \frac{2 \,\sigma_{\pi N}}{m_N \left(1+\frac{m_d}{m_u}\right)\left(1+\frac{B_d^{(N)}}{B_u^{(N)}}\right)}, \nonumber \\
f_{d}^{(N)} &=& \frac{m_dB_d^{(N)}}{m_N} = \frac{2 \,\sigma_{\pi N}}{m_N \left(1+\frac{m_u}{m_d}\right)\left(1+\frac{B_u^{(N)}}{B_d^{(N)}}\right)}, \\ 
f_{s}^{(N)} &=& \frac{m_sB_s^{(N)}}{m_N} = \frac{\frac{m_s}{m_d}\, y \,\sigma_{\pi N}}{m_N \left(1+\frac{m_u}{m_d}\right)}. \nonumber
\end{eqnarray}
We take the light quark mass ratios to be $m_u/m_d = 0.553$ and $m_s/m_d = 18.9$~\cite{quarkmasses}, and adopt the default values
$\sigma_{\pi N} = 55$ MeV and $\sigma_0 = 35$ MeV from~\cite{micromegas}, leading to
\begin{equation}
\setlength\arraycolsep{0.1em}
 \begin{array}{ccccc}
f_{u}^{(p)} &=& 0.023, \,\,\,\,f_{d}^{(p)} &= 0.033, \,\,\,\,f_{s}^{(p)} &= 0.26, \\
f_{u}^{(n)} &=& 0.018, \,\,\,\,f_{d}^{(n)} &= 0.042, \,\,\,\,f_{s}^{(n)} &= 0.26.
\end{array}
\end{equation}

We note that there is significant uncertainty in the value of the pion-nucleon sigma term, as explored recently in~\cite{Ellis:2008hf}. It was found that varying $\sigma_{\pi N}$ from its minimal value, $\sigma_0$, to the $2\sigma$ upper bound of 80 MeV can result in a change in $\sigma_{SI}$ by as much as a factor of $\sim10$, depending somewhat on the location of the point in the CMSSM parameter space for which the calculation is carried out.  Similar effects would be observed in NUHM models.  Since we choose $\sigma_{\pi N}=55$ MeV, the exact values of $\sigma_{SI}$ reported here may therefore be systematically offset by a factor of a few.  We caution the reader to interpret any apparent exclusion with this in mind, and rather to focus on the broader trends in the following analysis.

\subsection{XENON constraints}

All models considered here are cosmologically viable, with $\Omega_{\na}h^2 < 0.12$, and respect the collider constraints detailed above.
Throughout the paper, if a particular model point has $\Omega_{\na}$ less than the central value of $\Omega_{CDM}$ given by Eq.~\ref{eq:relic},
we take the local WIMP density entering the XENON detector to be reduced by the fraction 
$\Omega_{\na}/\Omega_{CDM}$.
Effectively, we compute a normalized scattering cross section,
\begin{equation}
\sigma_{SI} \rightarrow \sigma_{SI} \times \frac{\Omega_{\na}}{\Omega_{CDM}} .
\label{eq:normalized}
\end{equation}
Since the count rate for low density LSPs in the detector is reduced, the bounds from XENON-100 on $\sigma_{SI}$ for these models are weaker and the discoverability in XENON-1T is reduced.  

\begin{figure}[h]
\includegraphics[width=80mm, height=80mm]{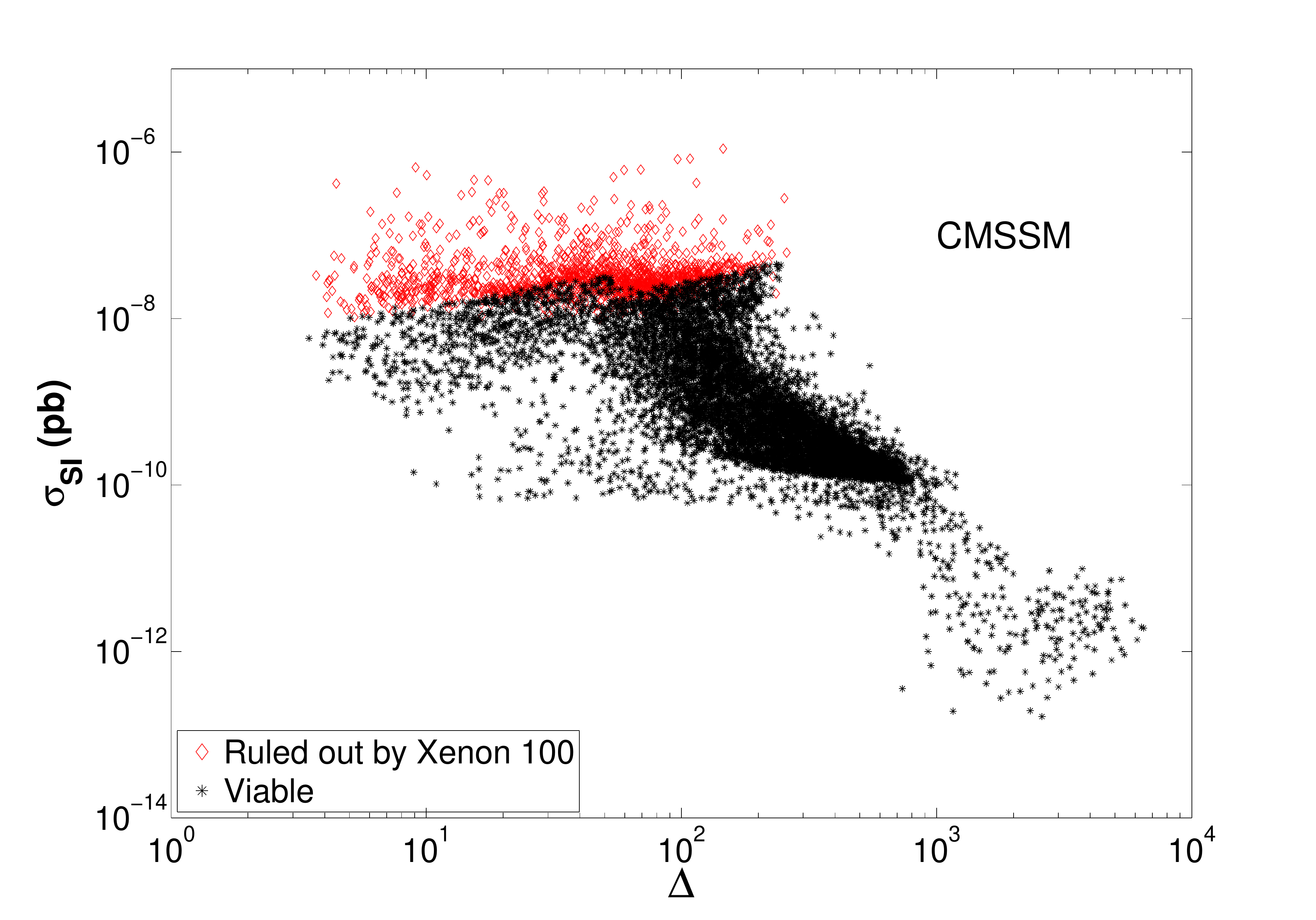}
\includegraphics[width=80mm, height=80mm]{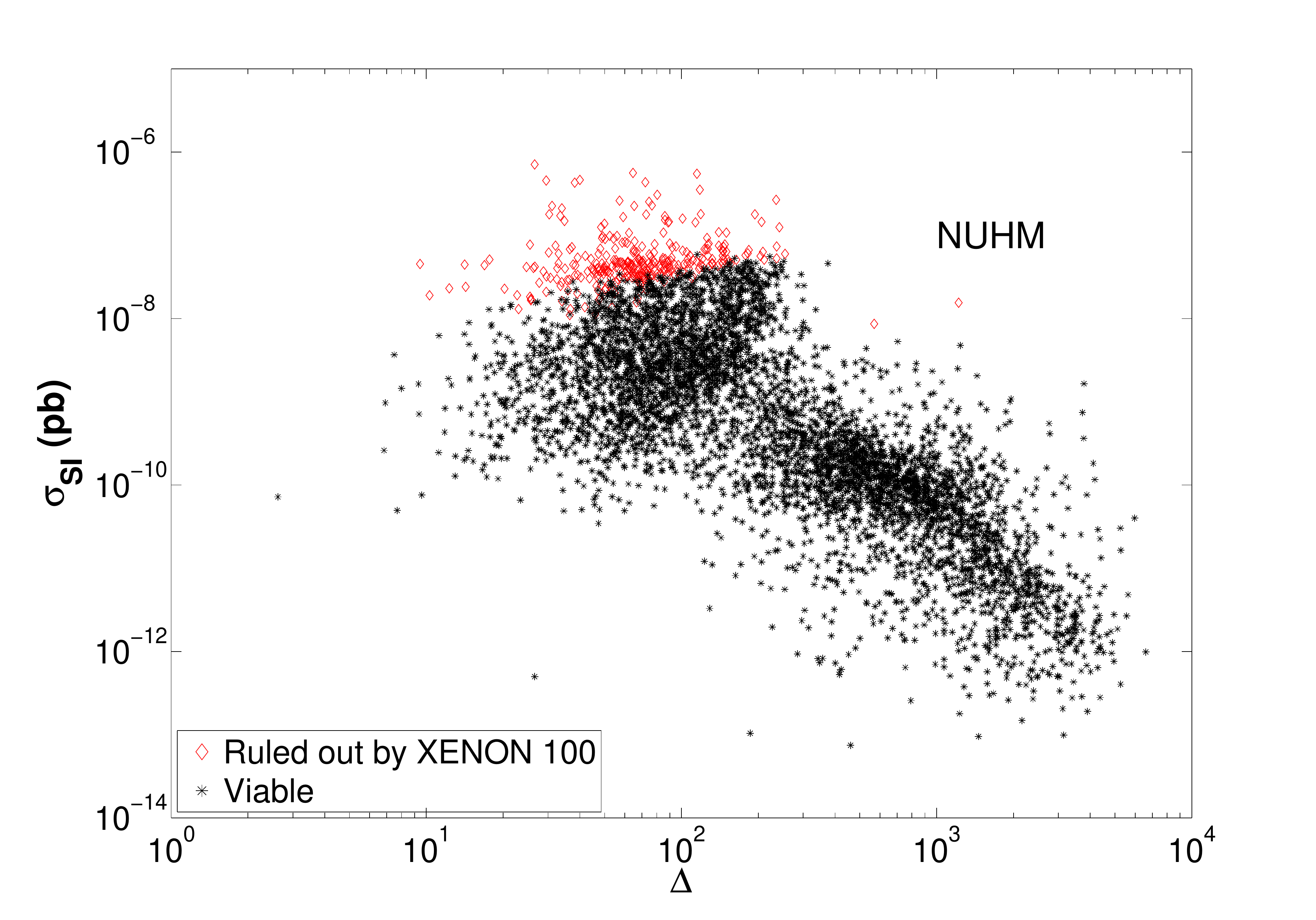}
\caption{Spin-independent neutralino-nucleon scattering cross section, $\sigma_{SI}$, as a function of fine-tuning parameter, $\Delta$, for the CMSSM (left) and
the NUHM (right).  Red points are ruled out by XENON-100 while black points are viable.
 A one-sided bound
$\Omega_{\na}h^2 < 0.12$ has been applied.}
\label{ruled_XENON}
\end{figure}

 \ref{ruled_XENON} illustrates the XENON-100 bounds on the total spin-indendepent neutralino-nucleon elastic scattering cross section, $\sigma_{SI}$, as a function of fine-tuning, $\Delta$, in the CMSSM and the NUHM.  Red points are ruled out by XENON-100 while black points are still viable.  
Clearly a far smaller fraction of the NUHM points are ruled out compared to CMSSM points. 
From the general downward slope of the points in the $(\Delta,\sigma_{SI})$ plane, it is evident that as $\Delta$ becomes large, 
the neutralino-nucleon elastic scattering cross sections tend to decrease in both the CMSSM and the NUHM.
This is related to the fact that large $\Delta$ implies large $\mu$, which, all other factors being fixed, would result in a more bino-like LSP.
Especially in the CMSSM, the least fine-tuned models tend to be the easiest to rule out, with the general trend that increasing sensitivity to $\sigma_{SI}$ will test increasingly fine-tuned models.

\begin{figure}[h]
\includegraphics[width=80mm, height=80mm]{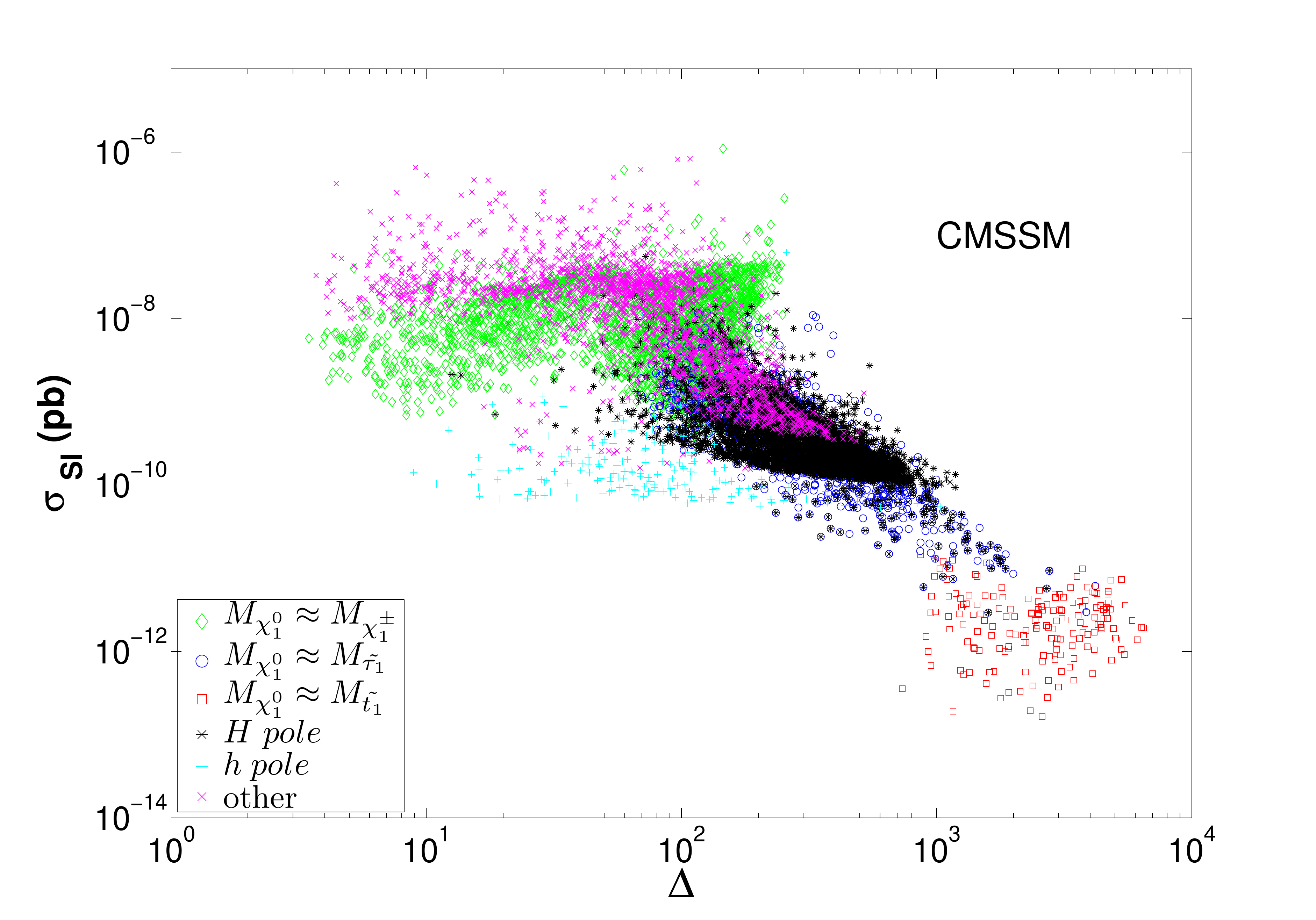}
\includegraphics[width=80mm, height=80mm]{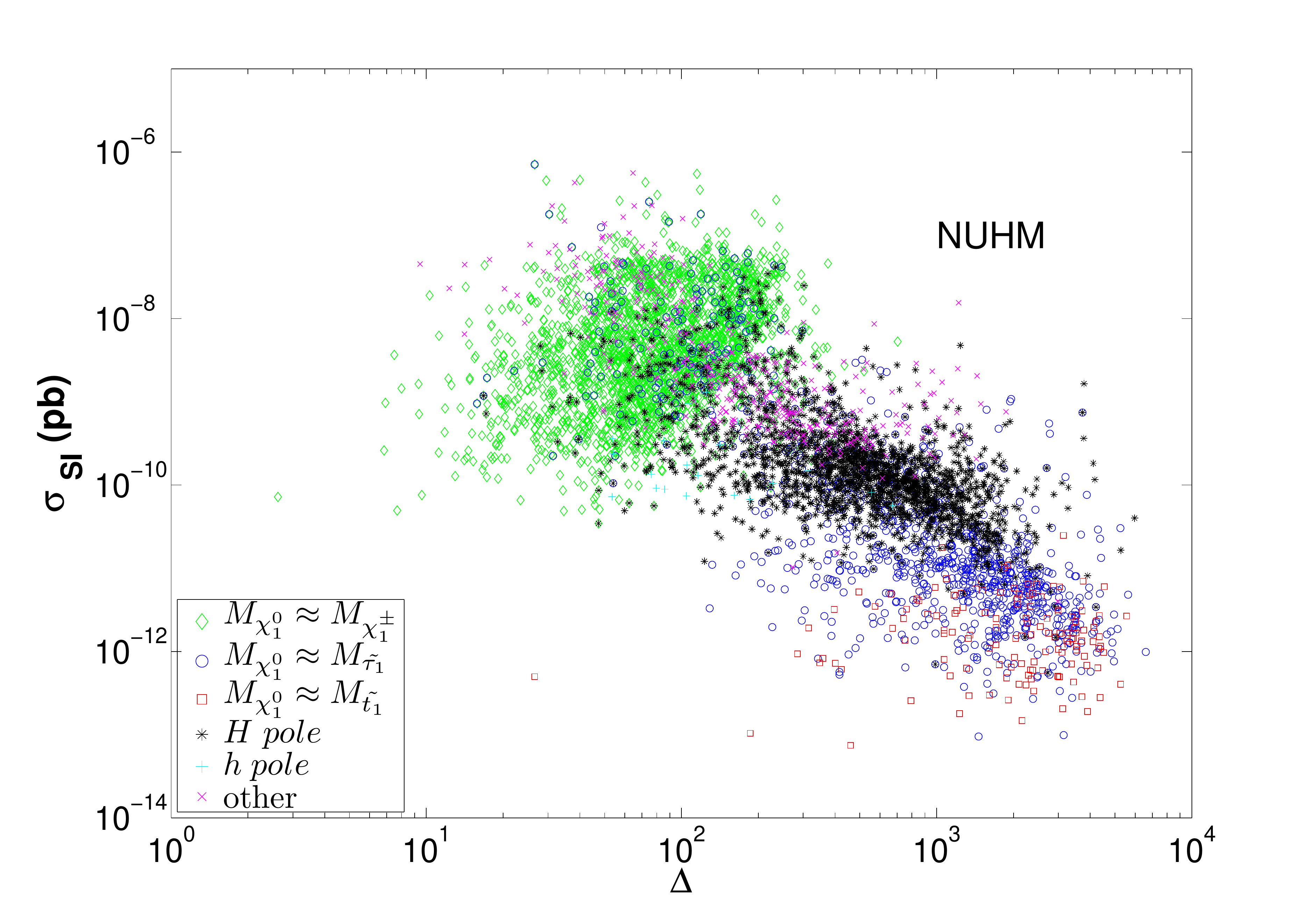}
\caption{Spin-independent neutralino-nucleon scattering cross section, $\sigma_{SI}$, as a function of fine-tuning parameter, $\Delta$, for the CMSSM (left) and
the NUHM (right).  Color-coding indicates SUSY mass hierarchy as described in the legend.
 A one-sided bound
$\Omega_{\na}h^2 < 0.12$ has been applied.}
\label{bloodybone}
\end{figure}

In the NUHM, the correlation between $\sigma_{SI}$ and fine-tuning does not hold as clearly.  Fig.~\ref{bloodybone} plots viable CMSSM and NUHM scenarios on the same axes, but illustrates the split into the various mass hierarchies as indicated.  In both the CMSSM and the NUHM, models with light charginos, as well as models that fall into the ``other'' category, are the least fine-tuned.  However, CMSSM scenarios with light charginos all have fairly large $\sigma_{SI}$ and will be probed by direct dark matter searches in the relatively near future (e.g.~XENON-1T), while in the NUHM, points with small fine-tuning and chargino NLSPs may be much more difficult to discover via direct dark matter searches. Given the additional freedom in the Higgs sector of the NUHM, it is perhaps surprising that the CMSSM and the NUHM exhibit as many similarities as they do.

\begin{figure}[h!]
\includegraphics[width=80mm, height=80mm]{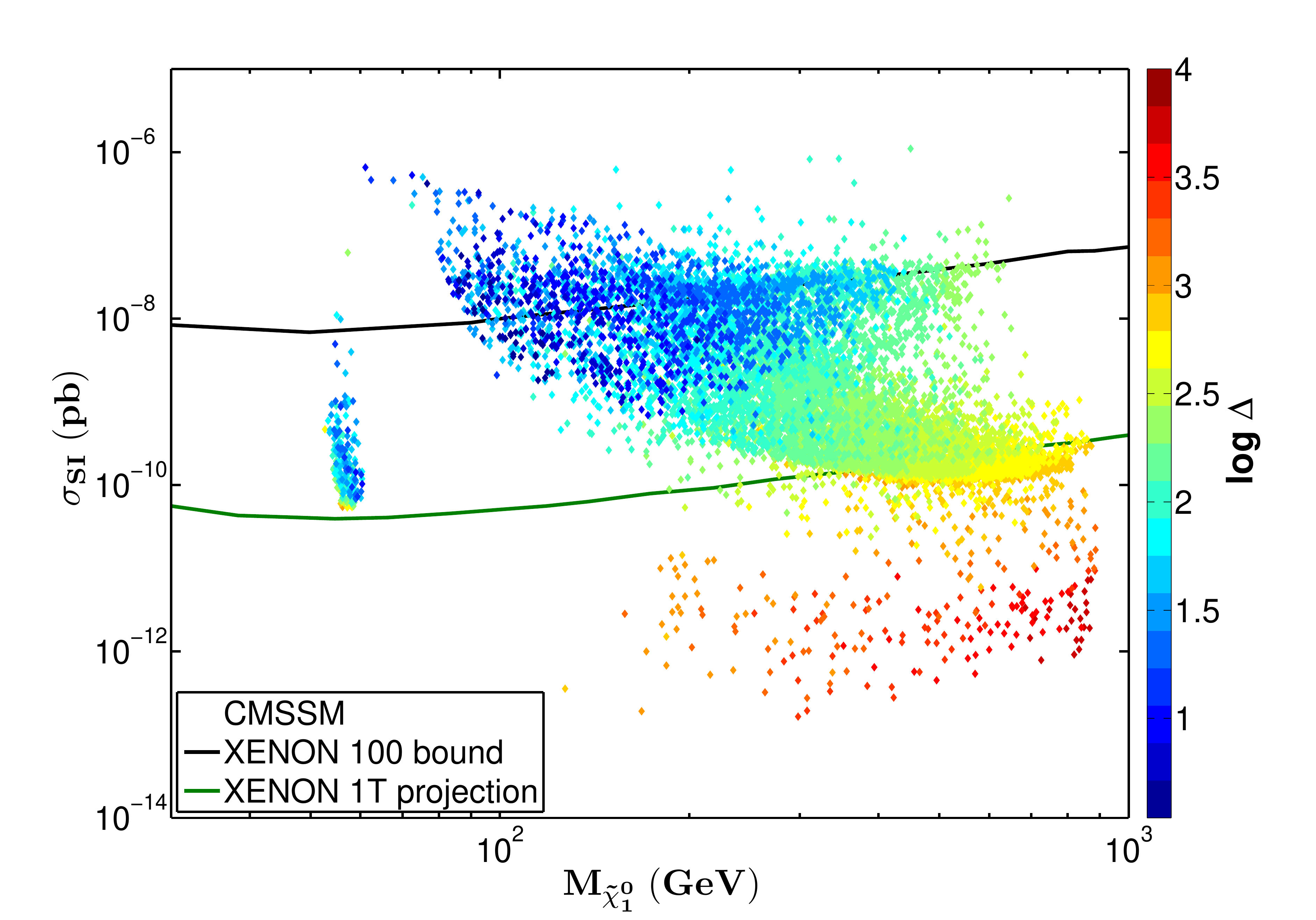}
\includegraphics[width=80mm, height=80mm]{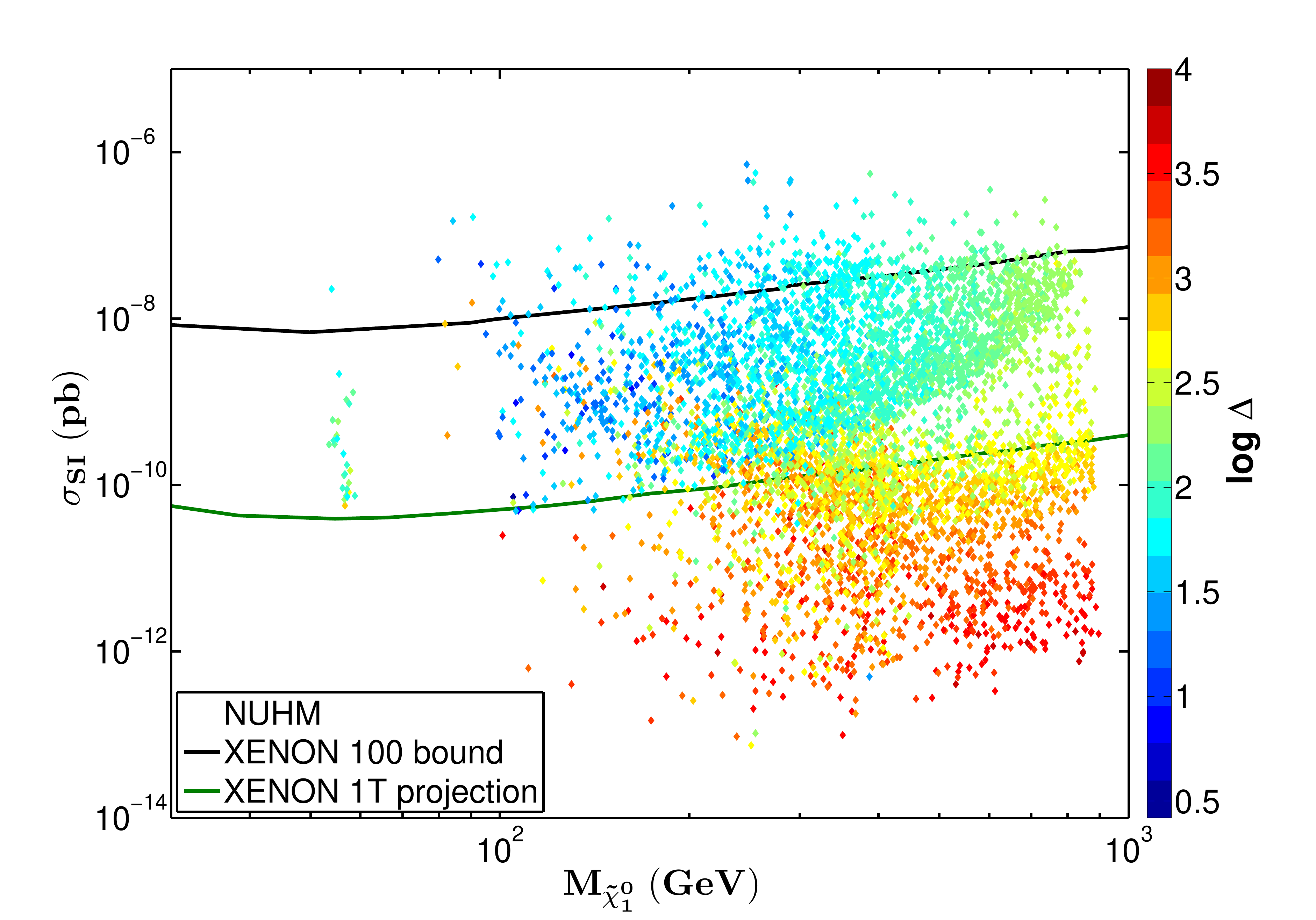}
\includegraphics[width=80mm, height=80mm]{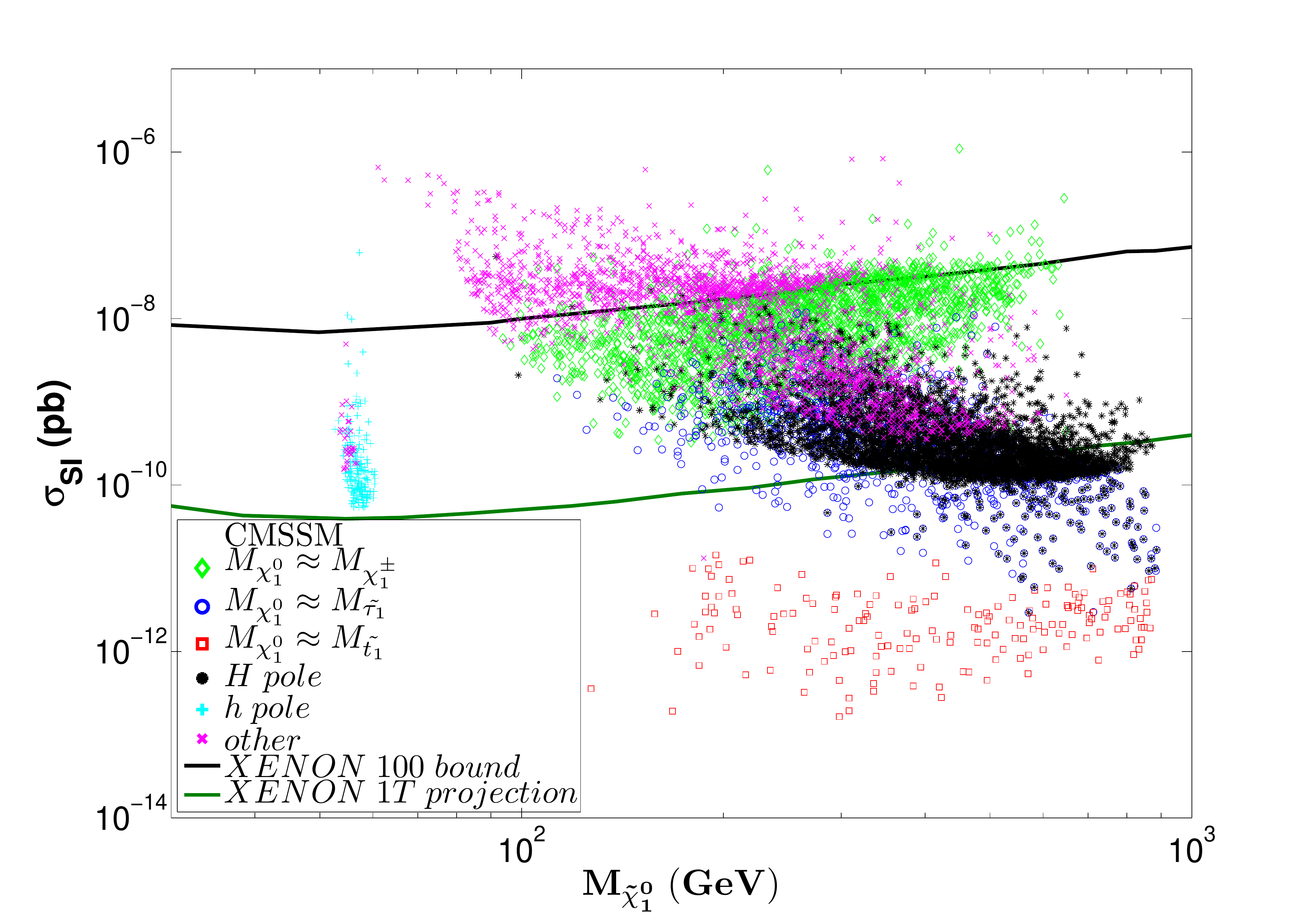}
\includegraphics[width=80mm, height=80mm]{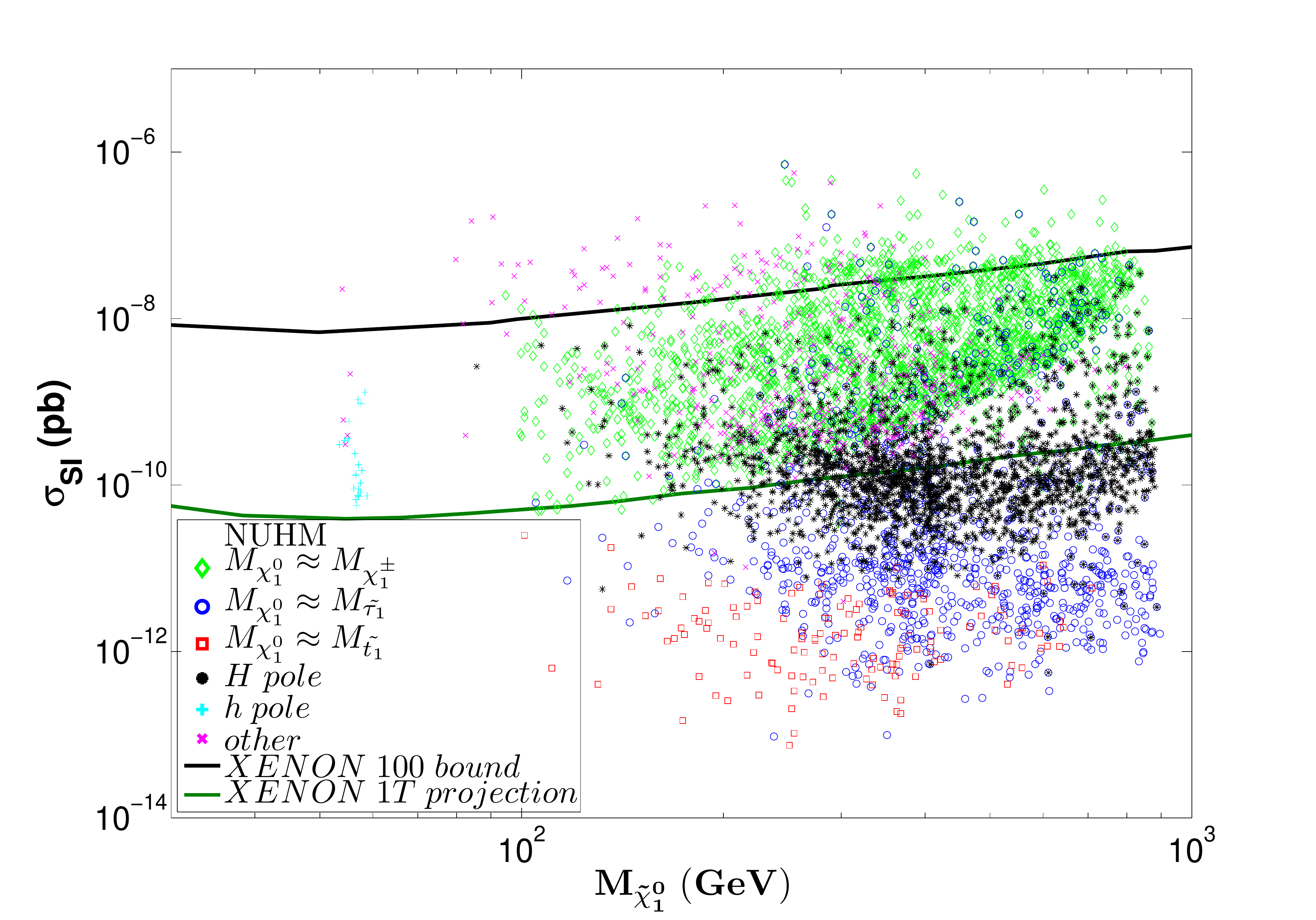}
\caption{Spin-independent neutralino-nucleon elastic scattering cross section, $\sigma_{SI}$, as a function of neutralino mass $M_{\na}$ for the CMSSM (left panels) and the NUHM (right panels). In the upper panels, model points are color-coded on a sliding scale according the the value of the fine-tuning parameter, $\Delta$, while in the lower panels, model points are color-coded by mass hierarchy as indicated in the legend. The current limit on $\sigma_{SI}$ from Xenon-100 is shown as the black curve, while the projected sensitivity of XENON-1T is represented by the green curve. 
A one-sided bound $\Omega_{\na}h^2 < 0.12$ has been applied.}\label{XENON}
\end{figure}

Further insight as to the differences between the CMSSM and the NUHM can be obtained by considering the $(m_{\na},\sigma_{SI})$ plane.
Fig.~\ref{XENON} illustrates the spin-independent neutralino-nucleon elastic scattering cross section as a function of neutralino mass for the CMSSM (left panels) and the NUHM (right panels).   The black (upper) and green (lower) curves in each panel represent the current upper limit on $\sigma_{SI}$ from XENON-100 and the projected sensitivity of XENON-1T, respectively. In the top panels, model points are color-coded on a sliding scale according to the value of the fine-tuning parameter, $\Delta$, while in the lower panels, model points are color-coded by mass hierarchy as indicated in the legend. 
From the top panels of Fig.~\ref{XENON}, it is evident that direct dark matter searches most easily test models with the least fine-tuning in EWSB (small $\Delta$), and probe progressively more fine-tuned models as experiments become more sensitive to $\sigma_{SI}$. In fact, the current limit from XENON-100 already excludes some of the least fine-tuned models. 

The relationship between $\sigma_{SI}$ and $\Delta$ can be understood by considering the role of $\mu$ in the determination of each quantity. As we have shown in Fig.~\ref{delta_mu}, the value of the fine-tuning parameter is strongly correlated with that of $\mu$, especially at large $\tan \beta$. In all cases, highly fine-tuned models have large $\mu$.  The composition of the lightest neutralino is also related to the value of $\mu$, i.e. for $\mu < M_1$ the neutralino LSP has a substantial higgsino component, while for $\mu \gg M_1$ it remains nearly entirely bino-like. Additionally, the spin-independent neutralino-nucleon elastic scattering cross section increases as the higgsino admixture increases.  So small $\Delta$ implies small $\mu$, which means the LSP is more likely to be substantially higgsino-like and therefore $\sigma_{SI}$ may be quite large.
Indeed, the top panels of Fig.~\ref{XENON} demonstrate that the least fine-tuned models are the ones most likely to be found in the next generation of direct detection experiments.  
We note, however, that as the LSP becomes purely higgsino, $\sigma_{SI}$ may again decrease: Since Higgs exchange is the dominant scattering process, and since Higgs exchange can occur only through gaugino-higgsino-Higgs couplings, a purely higgsino LSP would result in suppressed $\sigma_{SI}$, also. 

There is significantly more variation in the neutralino-nucleon elastic scattering cross section in the NUHM than in the CMSSM, especially for $m_{\na} \lesssim 150$ GeV or $m_{\na} \gtrsim 700$ GeV.  This, too, is a consequence of the additional freedom in the Higgs sector in the NUHM: Since $\mu$ is fixed by the electroweak vacuum conditions, which are related to the Higgs scalar masses, 
the LSP can be made Higgsino-like for nearly all choices of $M_{1/2}$ and $M_0$.  Furthermore, it is possible to maintain nearly the measured value of the relic density of neutralinos even if they are nearly completely higgsino-like.  This is not possible in the CMSSM, where dominantly higgsino LSPs have $\Omega_{\na}$ well below the WMAP-measured range for $\Omega_{CDM}$.  In both the CMSSM and the NUHM, if model points have $\Omega_{\na} \ll \Omega_{CDM}$, they appear in Fig.~\ref{XENON} as having significantly scaled $\sigma_{SI}$.  
For dominantly higgsino LSPs, the scaling is inevitable in the CMSSM, while the NUHM will contain CMSSM points (and others) that are significantly scaled as well as points for which no scaling is necessary.  Additionally, the LSP may be purely higgsino in the NUHM, and therefore have very low $\sigma_{SI}$, while this does not occur in the CMSSM.
The result is a larger range of effective scattering cross sections for the NUHM. 
However, a third and more dominant effect also stems from the additional freedom in the Higgs sector of the NUHM: The Higgs masses are not constrained by the choice of $M_0$ in the NUHM, so a larger range of Higgs masses are possible. Since $\sigma_{SI} \propto 1/m_H^4$ for scattering via Higgs exchange, there is a much larger range of possible scattering cross sections in the NUHM than in the CMSSM.  Higgs masses are bounded from below by collider constraints in all cases, so the amount by which the Higgs masses in NUHM scenarios can be smaller than those in the CMSSM is limited.  However, Higgs masses can be much larger in the NUHM than in the CMSSM, resulting in lower scattering cross sections. These findings are consistent with those presented in~\cite{nuhm1}.

\begin{figure}[h]
\includegraphics[width=80mm, height=80mm]{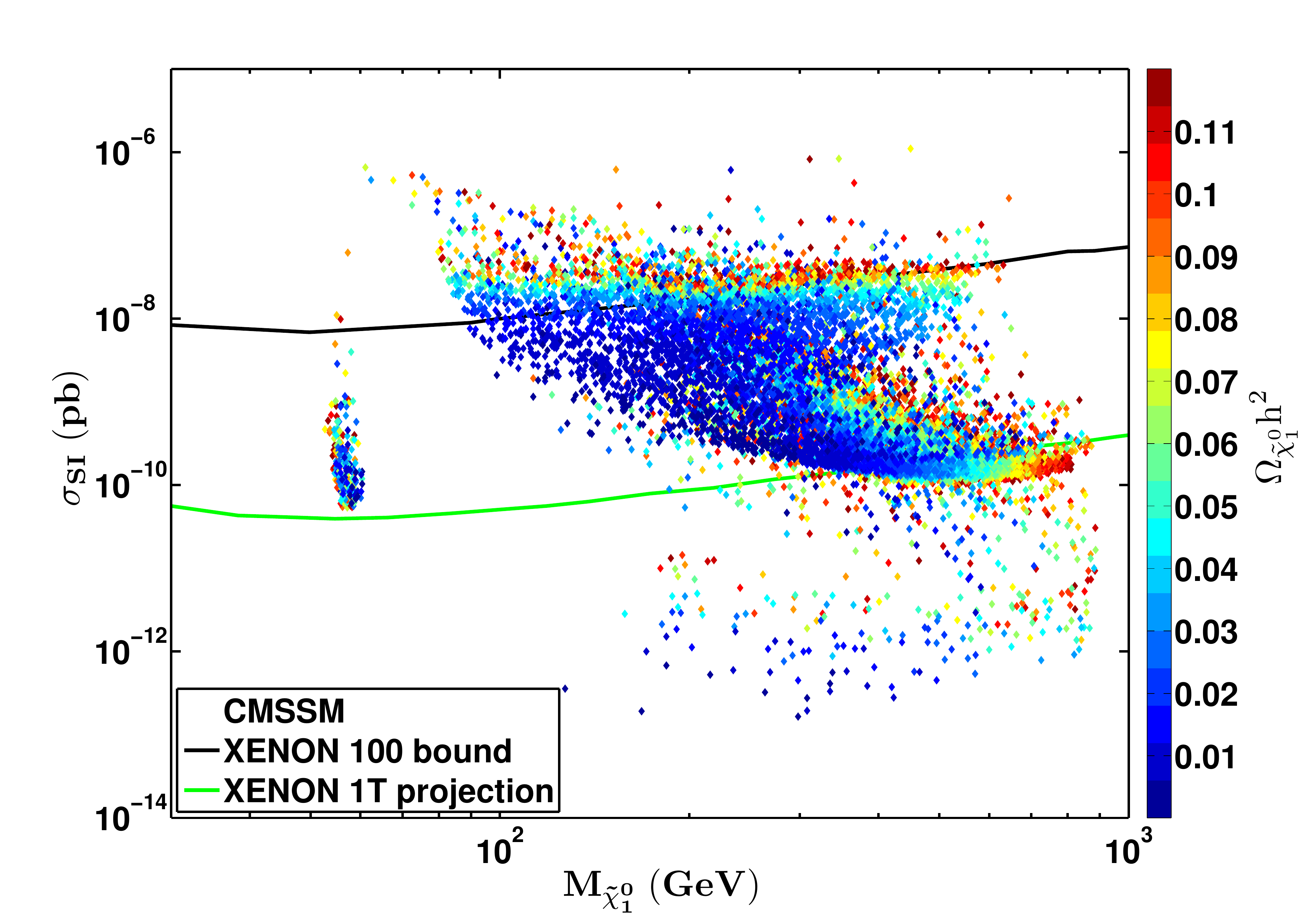}
\includegraphics[width=80mm, height=80mm]{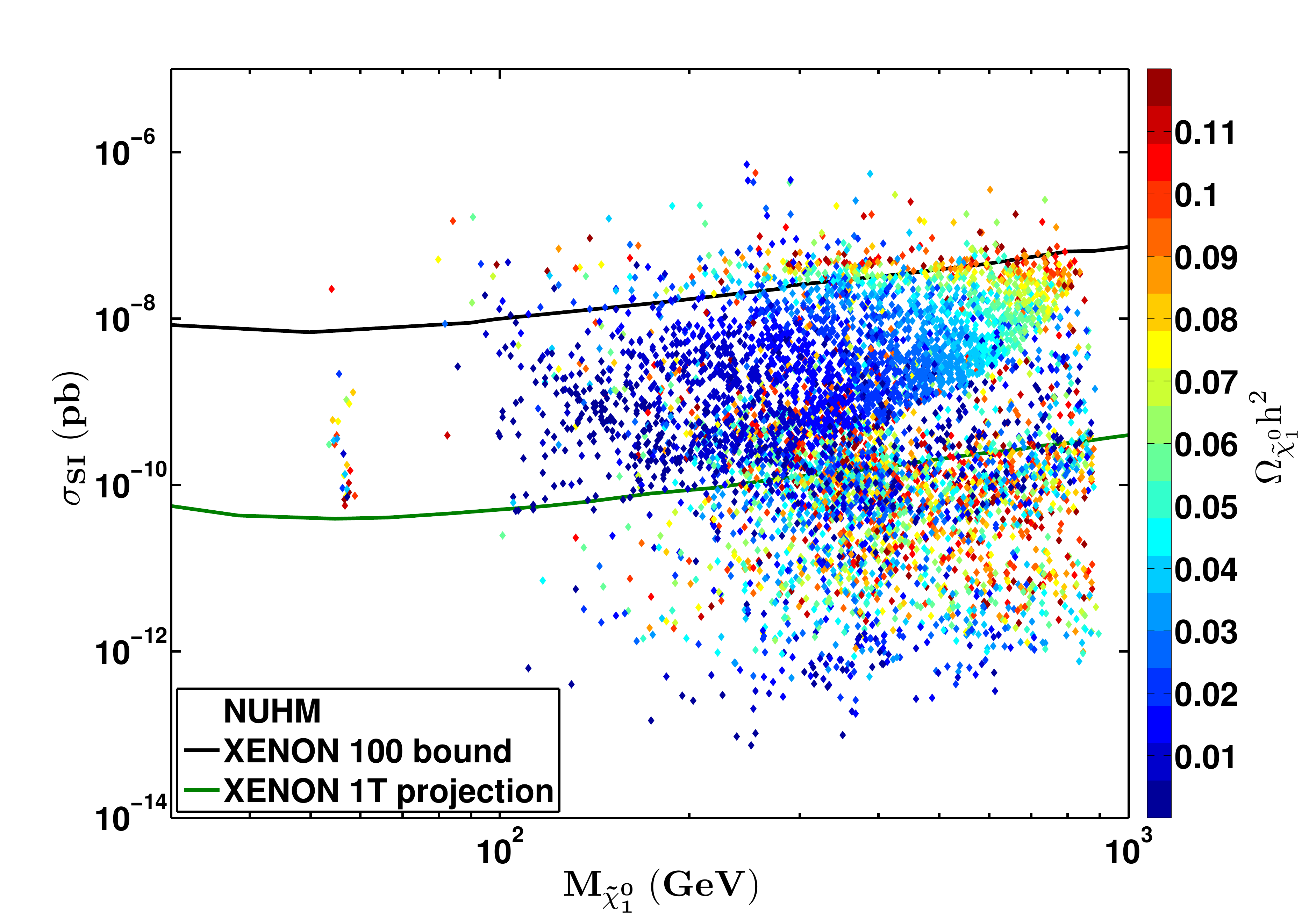}
\includegraphics[width=80mm, height=80mm]{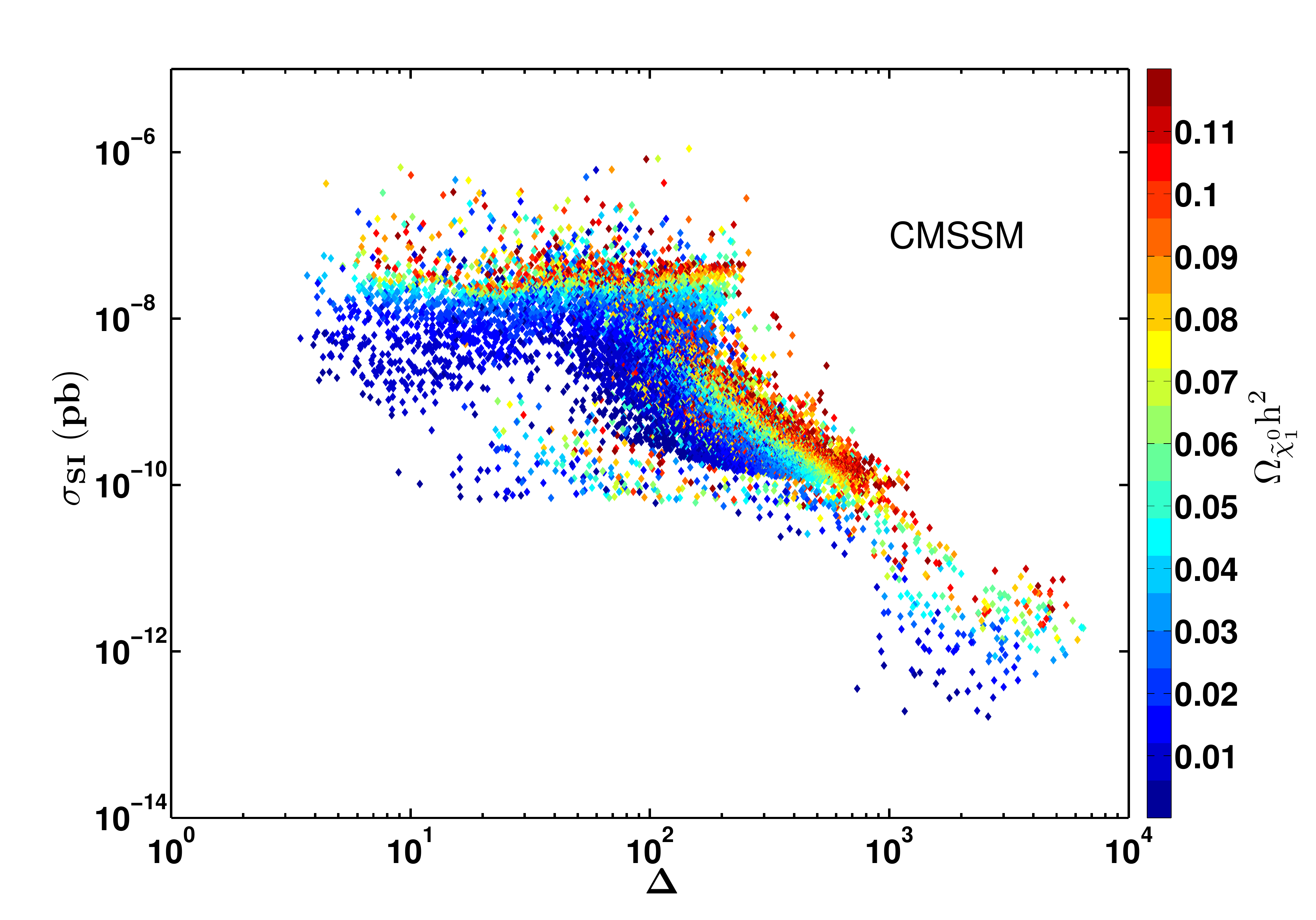}
\includegraphics[width=80mm, height=80mm]{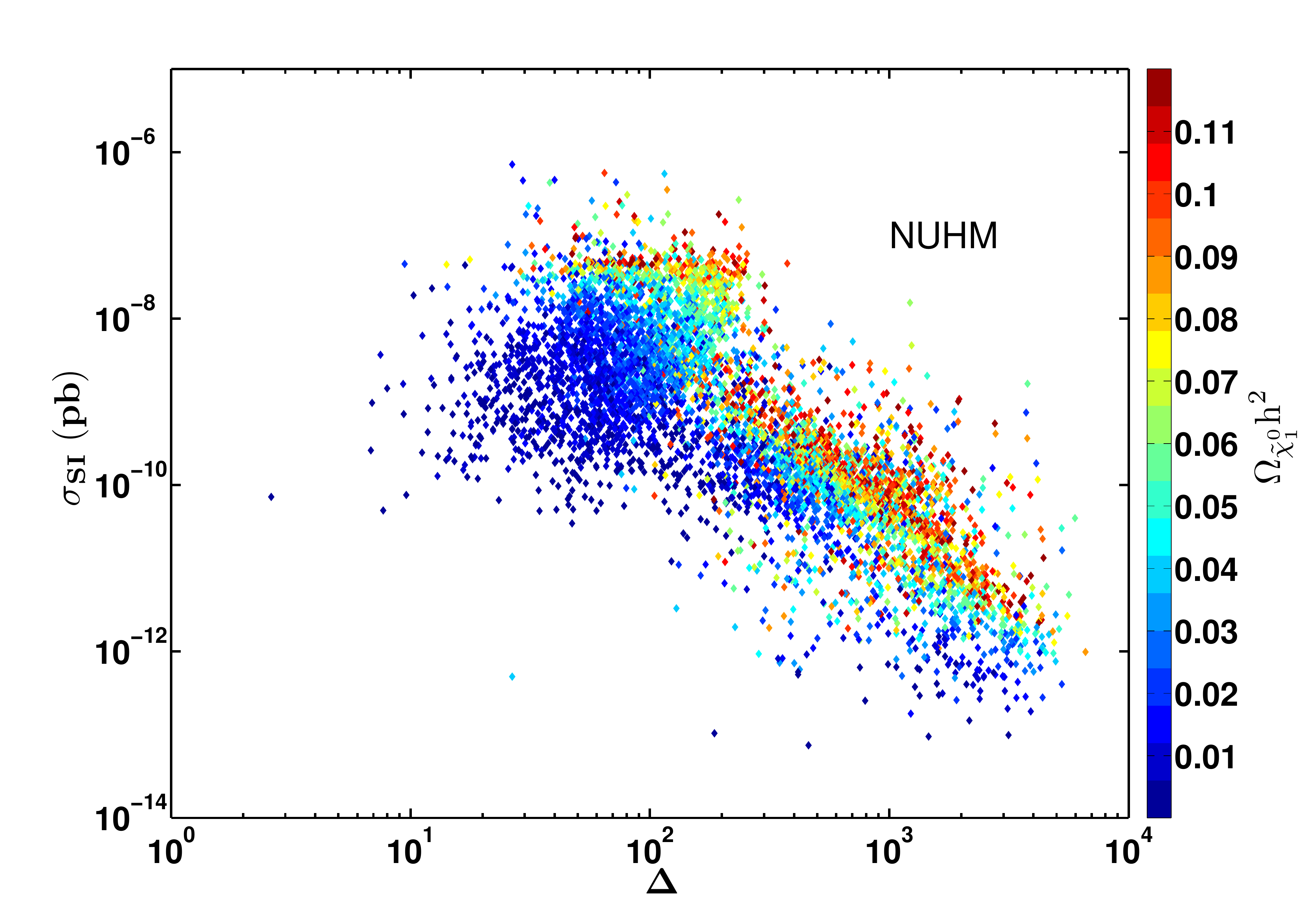}
\caption{In the top panels, the scattering cross section, $\sigma_{SI}$, is plotted against $m_{\na}$ with both the Xenon-100 bound and the projected sensitivity of the Xenon 1T experiment as shown. In the bottom panels, the scattering cross section is plotted against the fine-tuning parameter, $\Delta$.
Left panels are for the CMSSM; right
panels are for the NUHM.  Points are colored by the value $\Omega_{\na} h^2$ in each case. 
}\label{XENON_Omega}
\end{figure}

Thus far, our discussion of viable models has required only that the relic abundance of neutralinos not exceed the measured dark matter abundance.  In many cases, the abundance of neutralino dark matter is quite small, such that a secondary source of astrophysical cold dark matter is necessary. In the top panels of Fig.~\ref{XENON_Omega}, we show the $(m_{\na},\sigma_{SI})$ plane, color-coded to illustrate the resulting value of $\Omega_{\na} h^2$ for each model.  
Again, all points satisfy the upper bound of $\Omega_{\na} h^{2} < 0.12$, but the variation in $\Omega_{\na} h^{2}$ is clear.
In the CMSSM, points with $\Omega_{\na}h^2 \approx \Omega_{CDM}h^2$ tend to have larger cross sections than points with $\Omega_{\na}h^2 \ll \Omega_{CDM}h^2$.  This is somewhat expected, given the scaling of $\sigma_{SI}$ according to Eq.~\ref{eq:normalized}. In the NUHM, however, a correlation between $\Omega_{\na} h^{2}$ and $\sigma_{SI}$ is less obvious; only for $m_{\na} \gtrsim 300$ GeV and relatively large $\sigma_{SI}$ is it somewhat apparent in the upper right panel of Fig.~\ref{XENON_Omega}. 

In the lower panels of Fig.~\ref{XENON_Omega}, we show the $(\Delta,\sigma_{SI}
)$ plane, again for the CMSSM (left) and NUHM (right), with the same color-coding for $\Omega_{\na}h^2$ as in the top panels. When displayed this way, the effect of scaling the scattering cross section by $\Omega_{\na}/\Omega_{CDM}$ is more clear: For any value of $\Delta$ (i.e. some small range of values of $\mu$), the largest cross sections tend to come from points with approximately the right relic abundance of neutralino dark matter, while points for which the abundance of neutralinos is far below $\Omega_{CDM}$ tend to have smaller effective $\sigma_{SI}$ after the scaling. We point out, however, that there are several scenarios in both the CMSSM and the NUHM where $\Omega_{\na} = \Omega_{CDM}$ but $\sigma_{SI}<10^{-10}$ pb.  In the CMSSM, the points with the lowest $\sigma_{SI}$ typically have very large fine-tuning of $\Delta \gtrsim 10^3$, while in the NUHM, there are many very low $\sigma_{SI}$ scenarios for $\Delta$ as small as $\sim 200$.     

\begin{figure}[h]
\includegraphics[width=80mm, height=80mm]{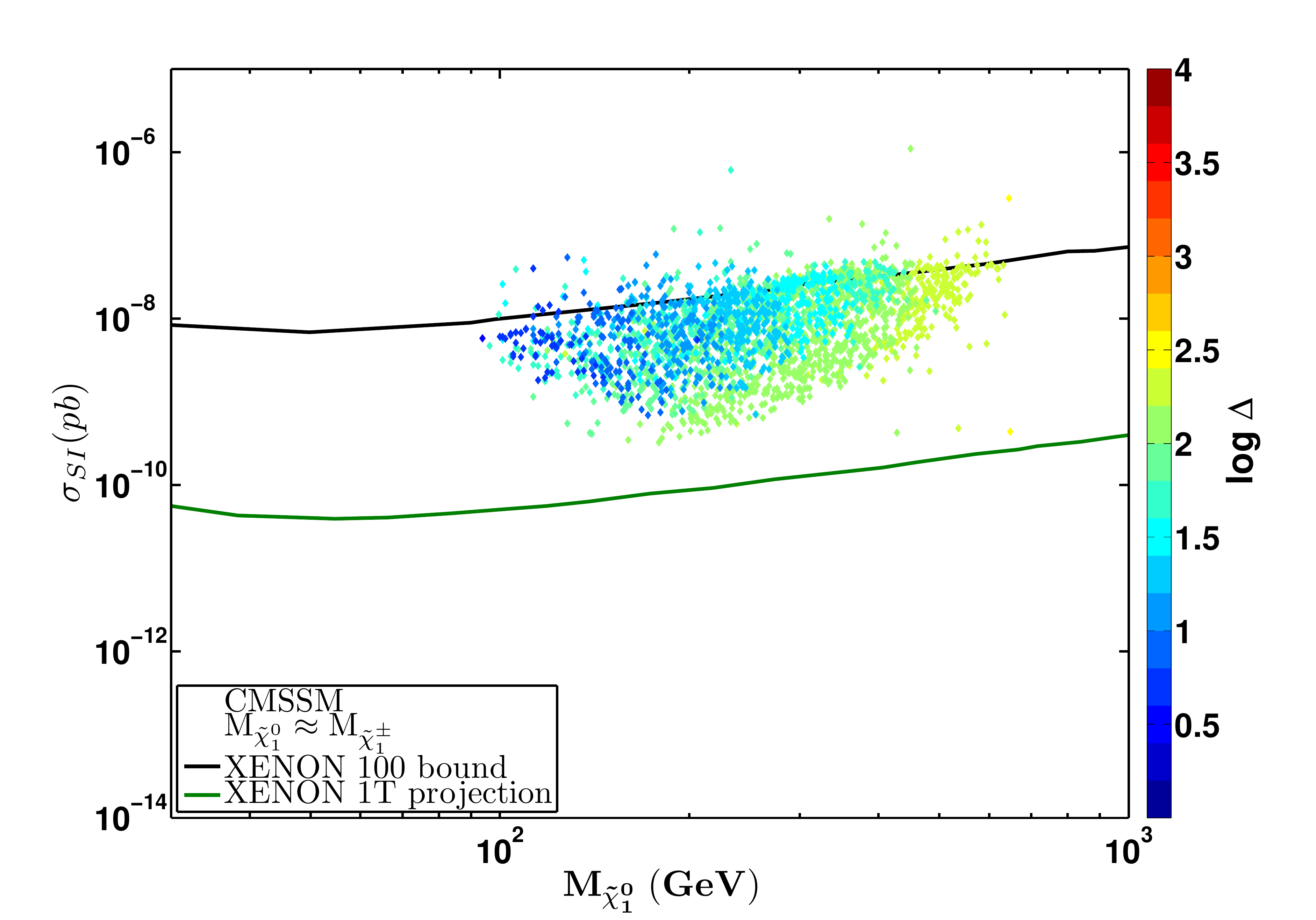}
\includegraphics[width=80mm, height=80mm]{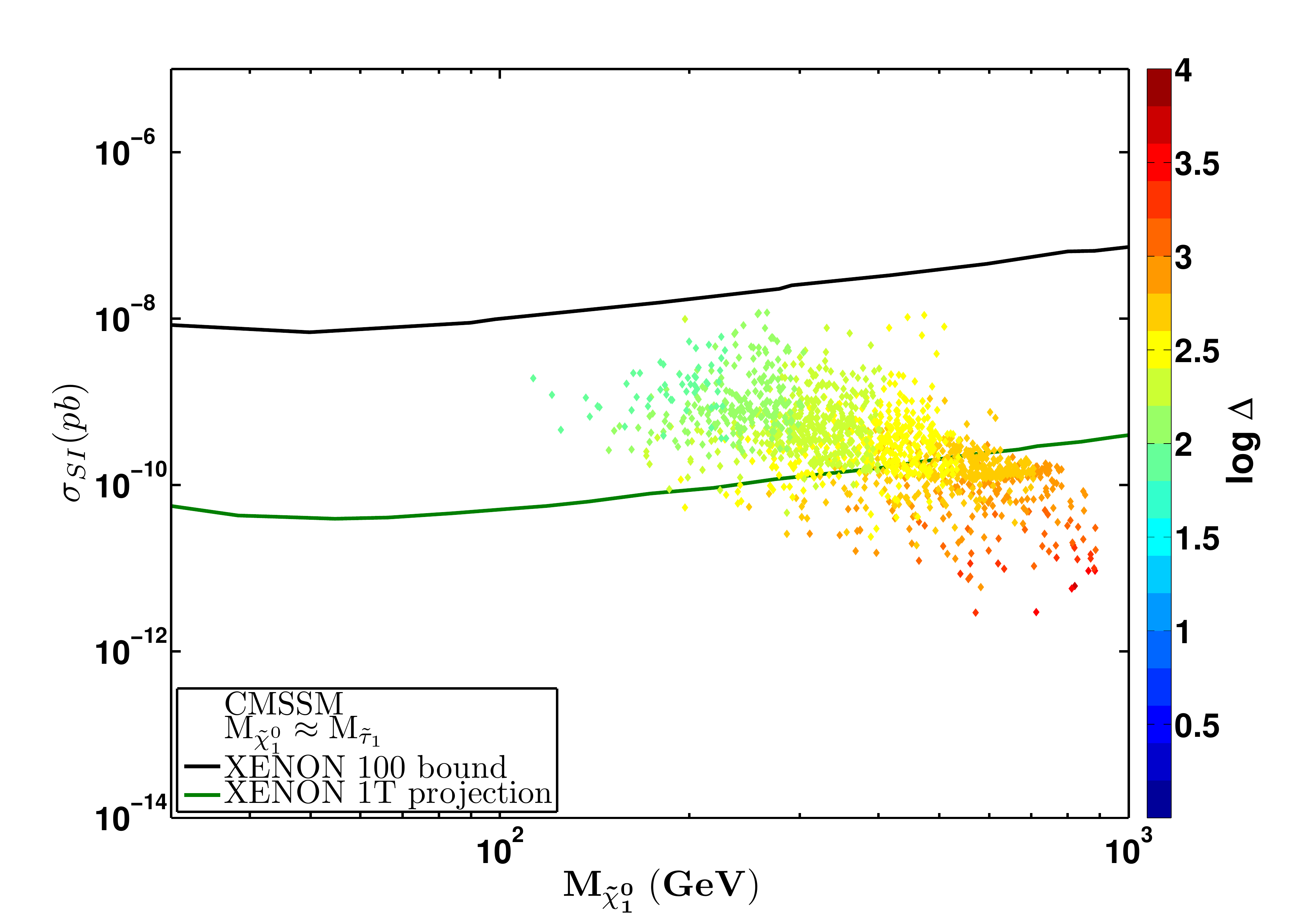}
\includegraphics[width=80mm, height=80mm]{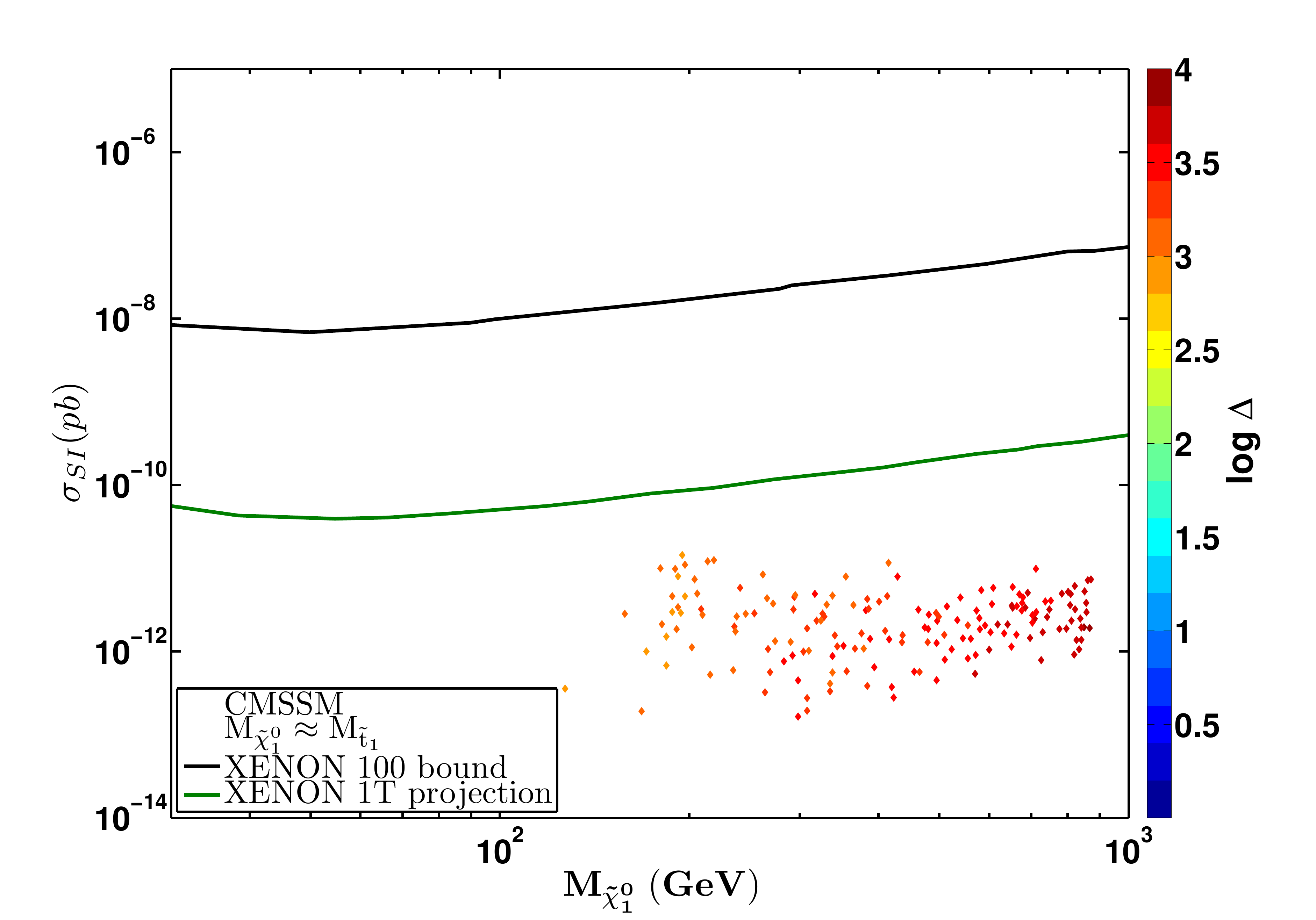}
\includegraphics[width=80mm, height=80mm]{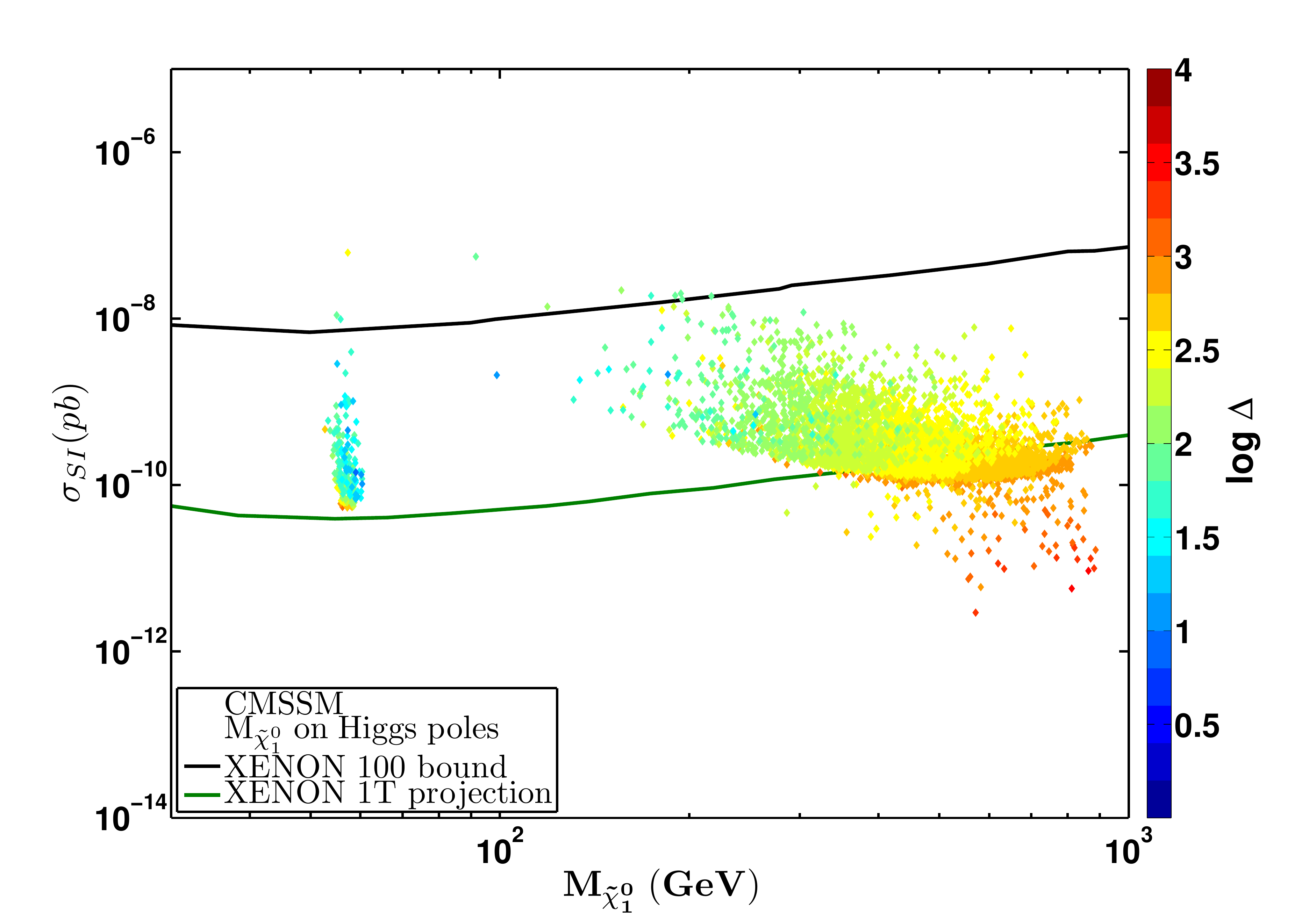}
\caption{Scattering cross-section $\sigma_{SI}$ is plotted against $M_{\na}$ for the CMSSM with the XENON 100 bound and projected XENON 1T sensitivity shown.
Models are split by SUSY mass-hierarchy.}
\label{msugra-split}
\end{figure}

\begin{figure}[h]
\includegraphics[width=80mm, height=80mm]{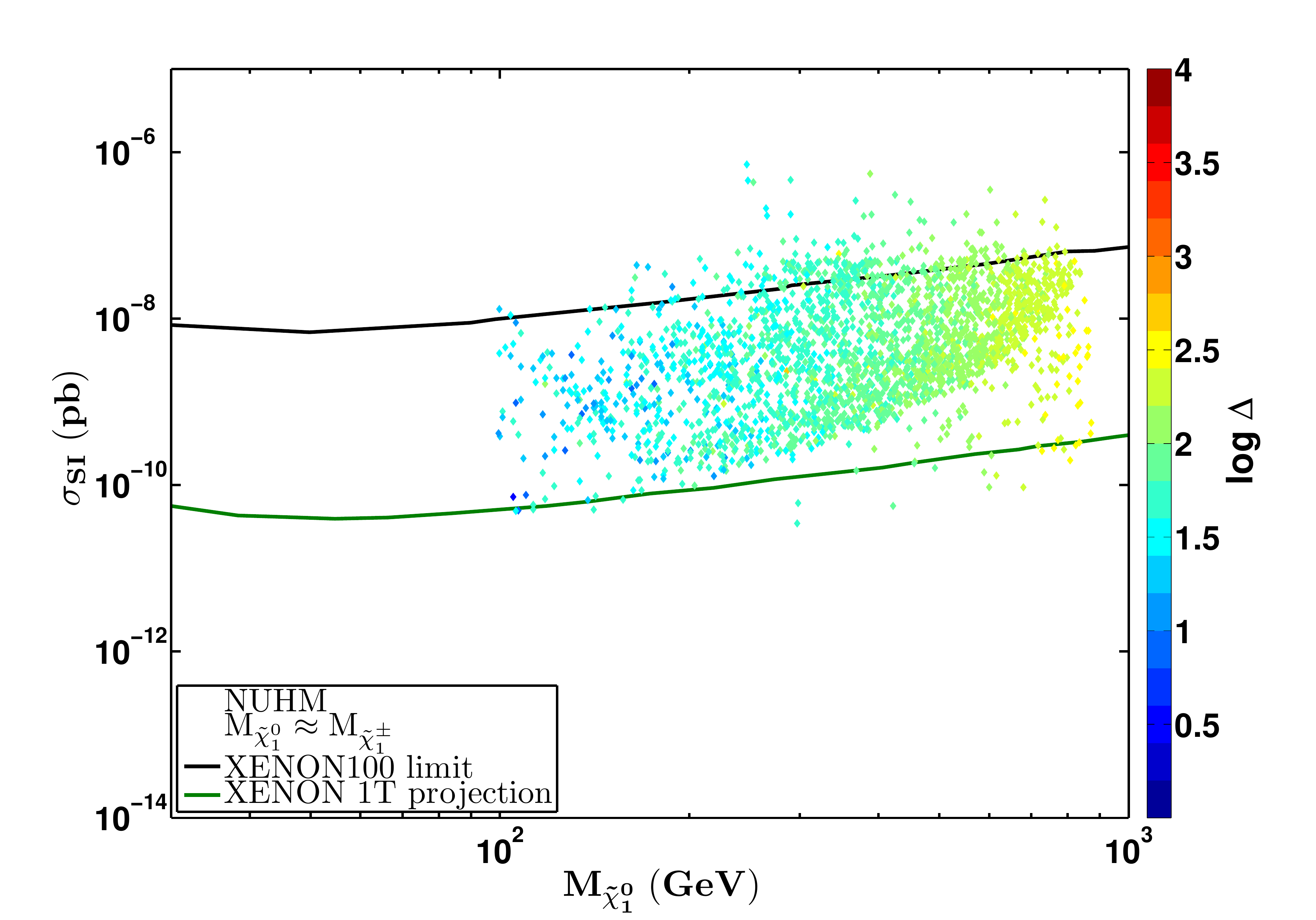}
\includegraphics[width=80mm, height=80mm]{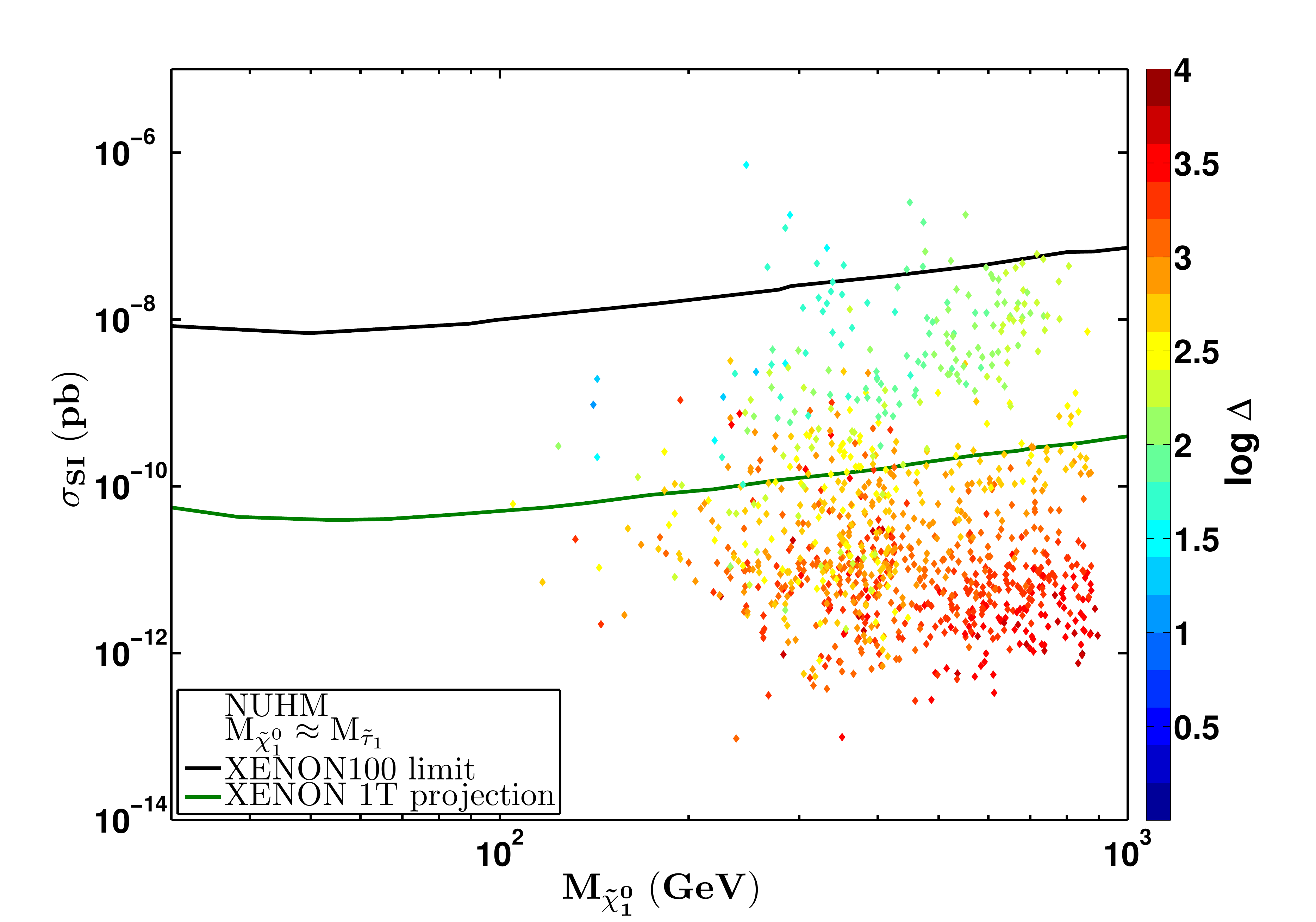}
\includegraphics[width=80mm, height=80mm]{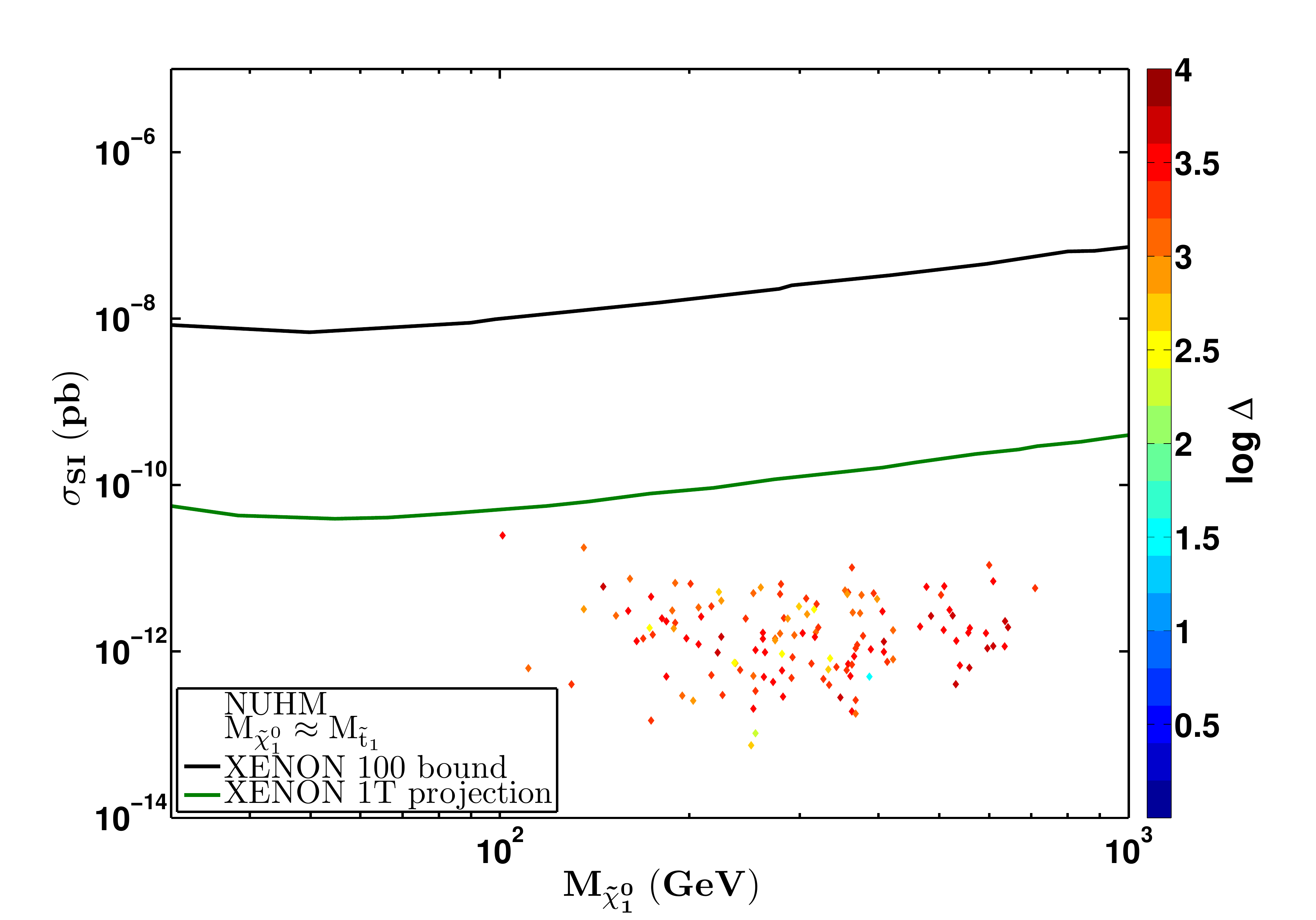}
\includegraphics[width=80mm, height=80mm]{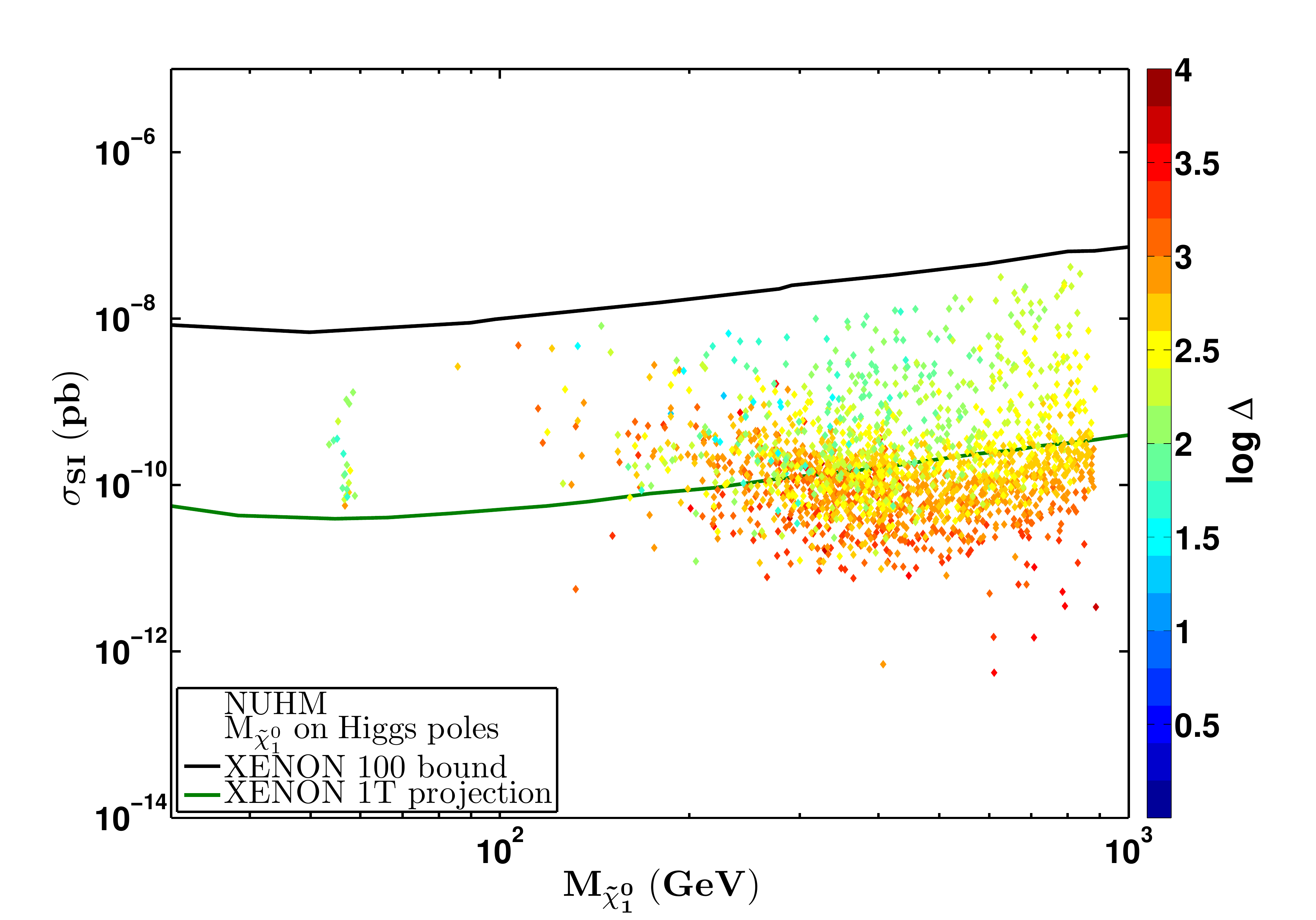}
\caption{Scattering cross-section $\sigma_{SI}$ is plotted against $M_{\na}$ for the NUHM with the XENON 100 bound and projected XENON 1T sensitivity shown.
Models are split by SUSY mass-hierarchy.}
\label{nuh-split}
\end{figure}

Returning to the question of the relationship between mass heirarchy and fine-tuning, 
Figs.~\ref{msugra-split} and \ref{nuh-split} show the $(m_{\na},\sigma_{SI})$ plane, with the current limit on $\sigma_{SI}$ from XENON-100 and the projected sensitivity of XENON-1T, for a variety of subsets of our CMSSM and NUHM parameter spaces chosen by mass hierarchy as described previously. In Fig.~\ref{msugra-split}, the CMSSM is explored, while in Fig.~\ref{nuh-split}, the NUHM is explored.

{\it Models with $m_{\na} \approx m_{\cha}$:}
The top-left panel of Figs.~\ref{msugra-split} and \ref{nuh-split} show the $(m_{\na}, \sigma_{SI})$ plane for the subset of CMSSM and NUHM scenarios in which the lighter chargino is nearly degenerate with the neutralino.
In these models, the neutralino
has a significant higgsino component:
If $\mu \gg M_1,M_2$, then $m_{\na}$ would be set by $M_1$ and $m_{\cha}$ would be set by $M_2$. However, in both the CMSSM and the NUHM, the ratio $M_1:M_2 \approx 1:2$ at the weak scale, so the lightest chargino would be about twice as massive as the lightest neutralino (in this case, co-annihilation of $\na$ with $\cha$ would not be possible). 
The models shown in the upper left panels of Figs.~\ref{msugra-split} and \ref{nuh-split} have $\mu \lesssim M_1$, and therefore a significantly higgsino-like LSP and lighter chargino.

In both the CMSSM and the NUHM, the value of the fine-tuning parameter $\Delta$ increases somewhat with WIMP mass: Since the neutralino LSP has a significant higgsino component, its mass is therefore related to $\mu$, which is in turn related to $\Delta$.  Another consequence of requiring $M_1>\mu$ is that the SU(3) gaugino, the gluino, also must be quite heavy, such that they may be more difficult to discover at the LHC.

As $\mu$ is relatively small for $m_{\na} \approx m_{\cha}$, these models typically have low fine-tuning and are among the most accessible at direct detection experiments, possessing the relatively large scattering cross sections associated with mixed bino-higgsino LSPs.
As already noted above, in the CMSSM, some of these points have already been ruled out by XENON-100, and all are well within the sensitivity of XENON-1T, while in the NUHM, the range of possible $\sigma_{SI}$ extends below the reach of XENON-1T.  We remind the reader that the exact sensitivity of direct detection experiments to neutralino-nucleon scattering depends on the nuclear form factors as discussed in section~\ref{sec:formalism}, and we use the projected XENON-1T reach primarily as a guide to compare the prospects in the CMSSM and the NUHM.

{\it Models with $m_{\na} \approx m_{\sta}$:} The top-right panels of Figs.~\ref{msugra-split} and \ref{nuh-split} show the $(m_{\na}, \sigma_{SI})$ plane for the subset of CMSSM and NUHM scenarios in which the lighter stau is nearly degenerate with the lightest neutralino.
In both the CMSSM and the NUHM, the least fine-tuned models are the most accessible to direct detection experiments. Since these models are defined by $m_{\na} \approx m_{\sta}$, if the neutralino LSP is light, the lighter stau will also be quite light, and therefore may be easily accessible at the LHC. In the CMSSM, all cases with very light $m_{\na} \approx m_{\sta} \lesssim 180$ GeV would be accessible to XENON-1T, however this conclusion does not hold for the NUHM, where there is considerably more variation in both $\sigma_{SI}$ and $\Delta$.
In both the NUHM and the CMSSM, there are scenarios with heavy $\na$ and $\sta$ that would not be discovered by XENON-1T.

 {\it Models with $m_{\na} \approx m_{\ta}$:}
The bottom left panels of Figs.~\ref{msugra-split} and \ref{nuh-split} show the $(m_{\na}, \sigma_{SI})$ plane for the subset of CMSSM and NUHM scenarios in which
the lighter stop is nearly degenerate with the lightest neutralino.
Although it seems that the neutralino in this case will not be
discoverable even with XENON-1T in either the CMSSM or the NUHM, a low $\ta$ mass is easily detectable at the LHC.   However, one can see that almost all of the points are quite fine-tuned with $\Delta > 1000$. 

One can understand the required high fine-tuning in the following way.
In order to get $m_{\ta}$ to be low enough to be close to the LSP mass, the running of $m_{\ta}$ must
be accelerated; this can be achieved with a large value of $|A_t| > 1$ TeV.  These large values of $A_{t}$ also drive $m_{H_{u}}$ to be large and negative. 
One can see from Eq.~\ref{zmass} that in order for EWSB to produce the observed value of $m_Z$, in the CMSSM, a large value of $\mu$ is then required, corresponding to large fine-tuning, $\Delta$.  Thus CMSSM models with $m_{\na} \sim m_{\ta}$ are quite fine-tuned.  Because of the additional freedom in the Higgs sector in the NUHM, it is possible for NUHM points with $m_{\na}\approx m_{\sta}$ to have somewhat lower fine-tuning than the corresponding points in the CMSSM.
However, the fine-tuning is uncomfortably large in both the CMSSM and the NUHM for $m_{\na} \approx m_{\ta}$. 

 {\it Models at a Higgs pole:}
The lower right panels of Figs.~\ref{msugra-split} and \ref{nuh-split} show the $(m_{\na}, \sigma_{SI})$ plane for the subset of CMSSM and NUHM scenarios in which annihilations of the lightest neutralino are enhanced by the presence of a Higgs pole.

The light Higgs pole is defined as $m_{\na} \approx m_h/2 \sim$ 50-60 GeV.  
In these cases, both $M_1$ and $\mu$ must be small to generate such a light neutralino LSP.
Since we have assumed gaugino mass unification at the GUT scale, the entire gaugino sector must then have correspondingly low mass. 
In the CMSSM, since $\mu$ is necessarily small in this region of parameter space, the fine-tuning, $\Delta$ is also small. In the NUHM, $\mu$, and therefore $\Delta$, may be somewhat larger. 
This region in CMSSM was previously 
studied in~\cite{Feldman:2011me}.

The heavy Higgs pole is defined as $m_{\na} \sim m_A/2$, where annihilations of lightest neutralinos through s-channel $A$-exchange are enhanced.  Here, $m_{\na} \gtrsim 90$ GeV.
Again, because of the additional freedom in the Higgs sector in the NUHM, the parameter space for A-pole annihilations is larger than in the CMSSM, resulting in a larger range of $\sigma_{SI}$ in the NUHM than in the CMSSM.  We note that the CMSSM is a subset of the NUHM, so the points in Fig.~\ref{msugra-split} that are excluded by XENON-100 would also appear in Fig.~\ref{nuh-split} had the parameter space scan been adequately dense. In the CMSSM, A-pole points at lower $m_{\na}$ and with larger $\sigma_{SI}$, i.e.~the most accessible to direct dark matter searches, are the least fine-tuned. In the NUHM, that conclusion does not hold; points with $\Delta$ as small as a few $\times 10$ have cross sections that will not be probed even by XENON-1T.

\section{Neutralinos with Correct Relic Density}

To this point, we have enforced only an upper bound on the neutralino relic density, 
 $\Omega_{\na} h^{2} < 0.12$.  In this section, we make the further restriction that neutralinos provide the entire
 content of the dark matter of the Universe, i.e., $\Omega_{\na} = \Omega_{CDM}$ in Eq. (\ref{eq:relic}).
Clearly far fewer points remain, but there are still some interesting trends.  

\begin{figure}[h]
\includegraphics[width=80mm, height=80mm]{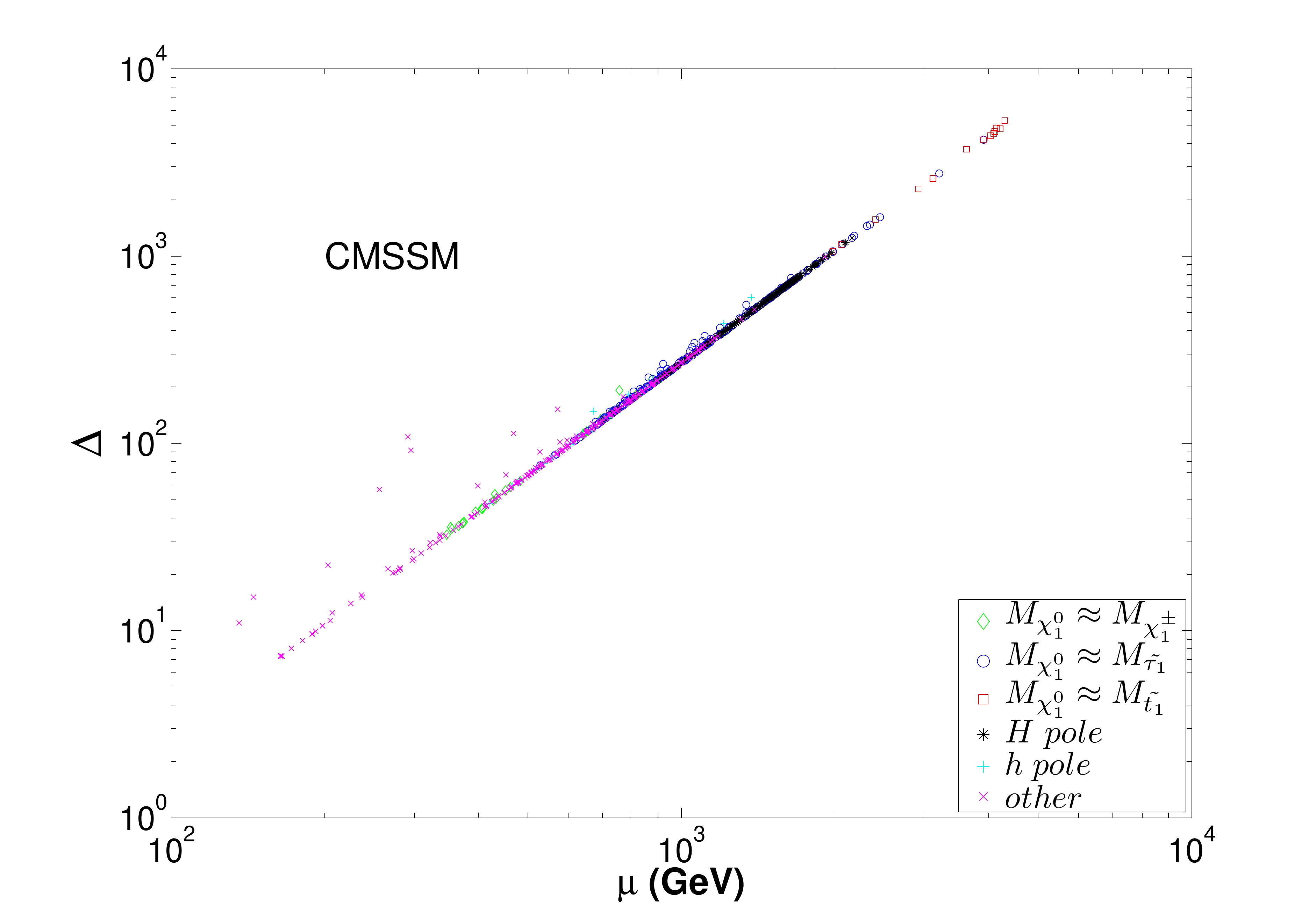}
\includegraphics[width=80mm, height=80mm]{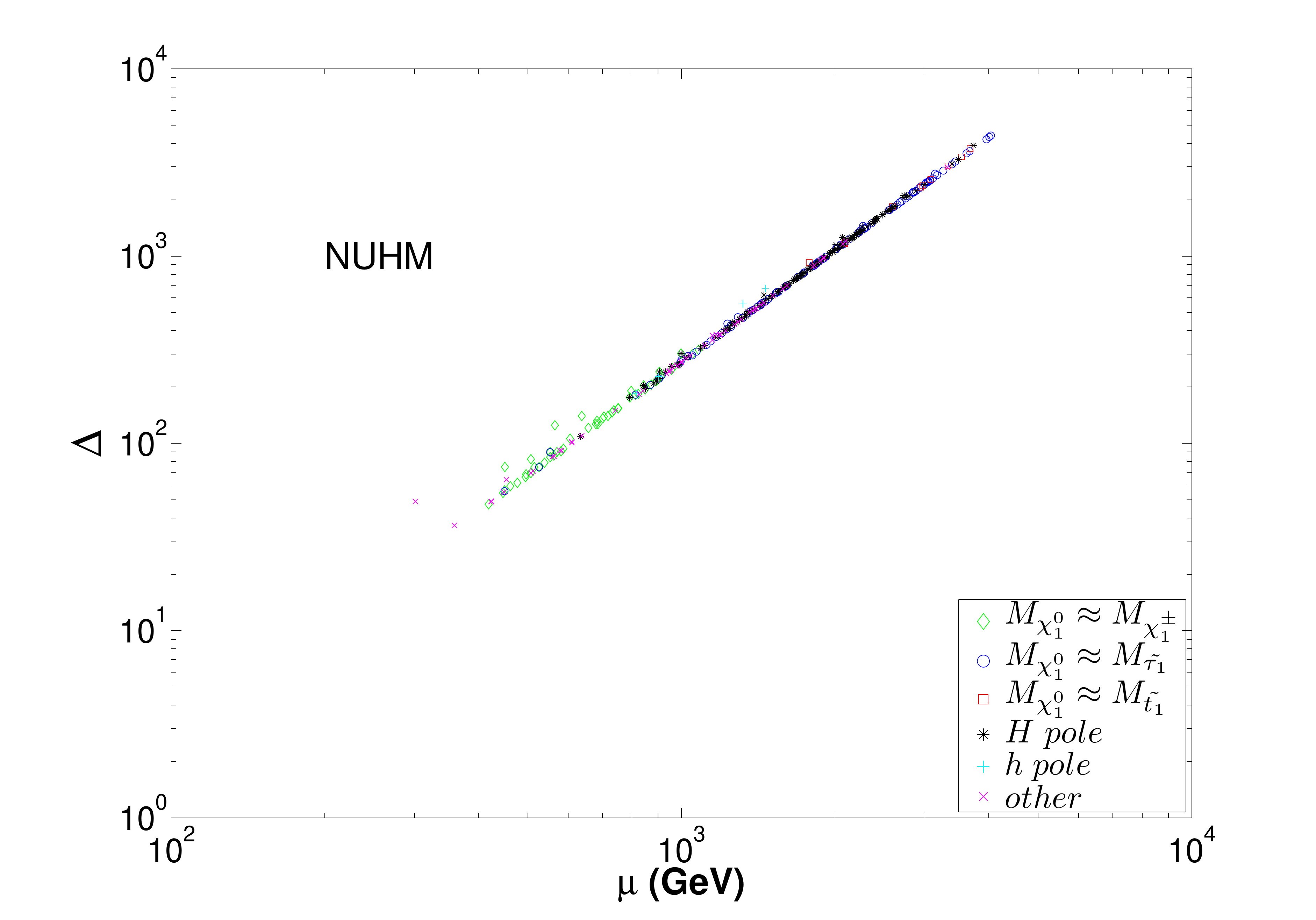}
\caption{Again, fine-tuning parametrized by $\Delta$ plotted against $\mu$. The scatter at low $\Delta$
disappears when the lower bound on $\Omega$ is enforced.}
\label{omega_delta_mu}
\end{figure}

With this additional constraint, the relation between $\Delta$ and $\mu$ is plotted in Fig.~\ref{omega_delta_mu}. 
The approximate relationship $\Delta \propto \mu^2$ still holds, and has far less scatter at low $\mu$ for the
following reason:  
As $\mu$ decreases below $M_1$, the lightest neutralino becomes increasingly higgsino-like, and less bino-like, resulting in a lower relic abundance of neutralinos. In many cases, this abundance is below the WMAP-measured dark matter range specified in Eq.~\ref{eq:relic}.
By comparison with Fig.~\ref{delta_mu}, many of the points at low $\mu$ have a neutralino abundance that is not sufficient to make up the dark matter, and are therefore absent from Fig.~\ref{omega_delta_mu}.

\begin{figure}[h]
\includegraphics[width=80mm, height=80mm]{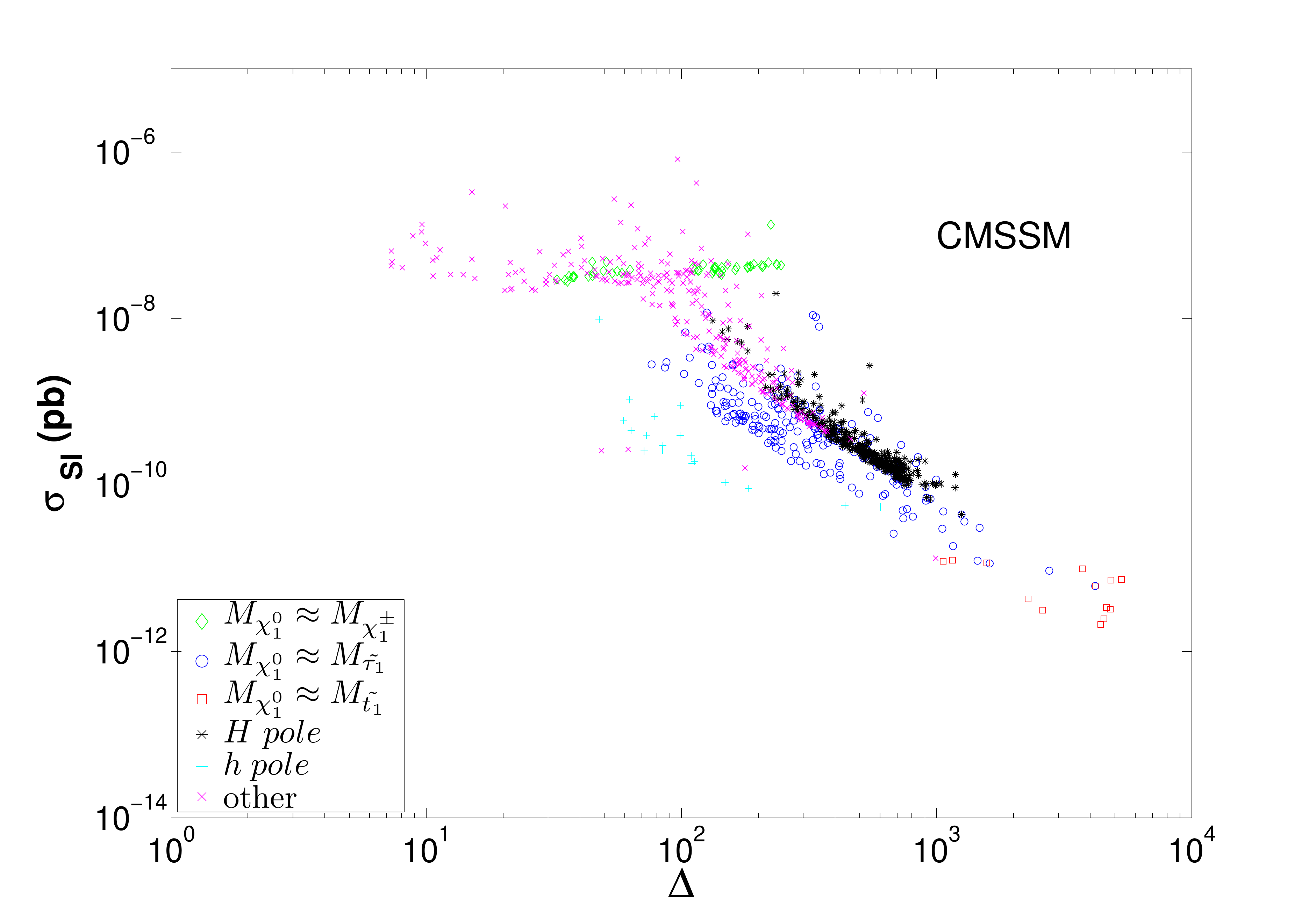}
\includegraphics[width=80mm, height=80mm]{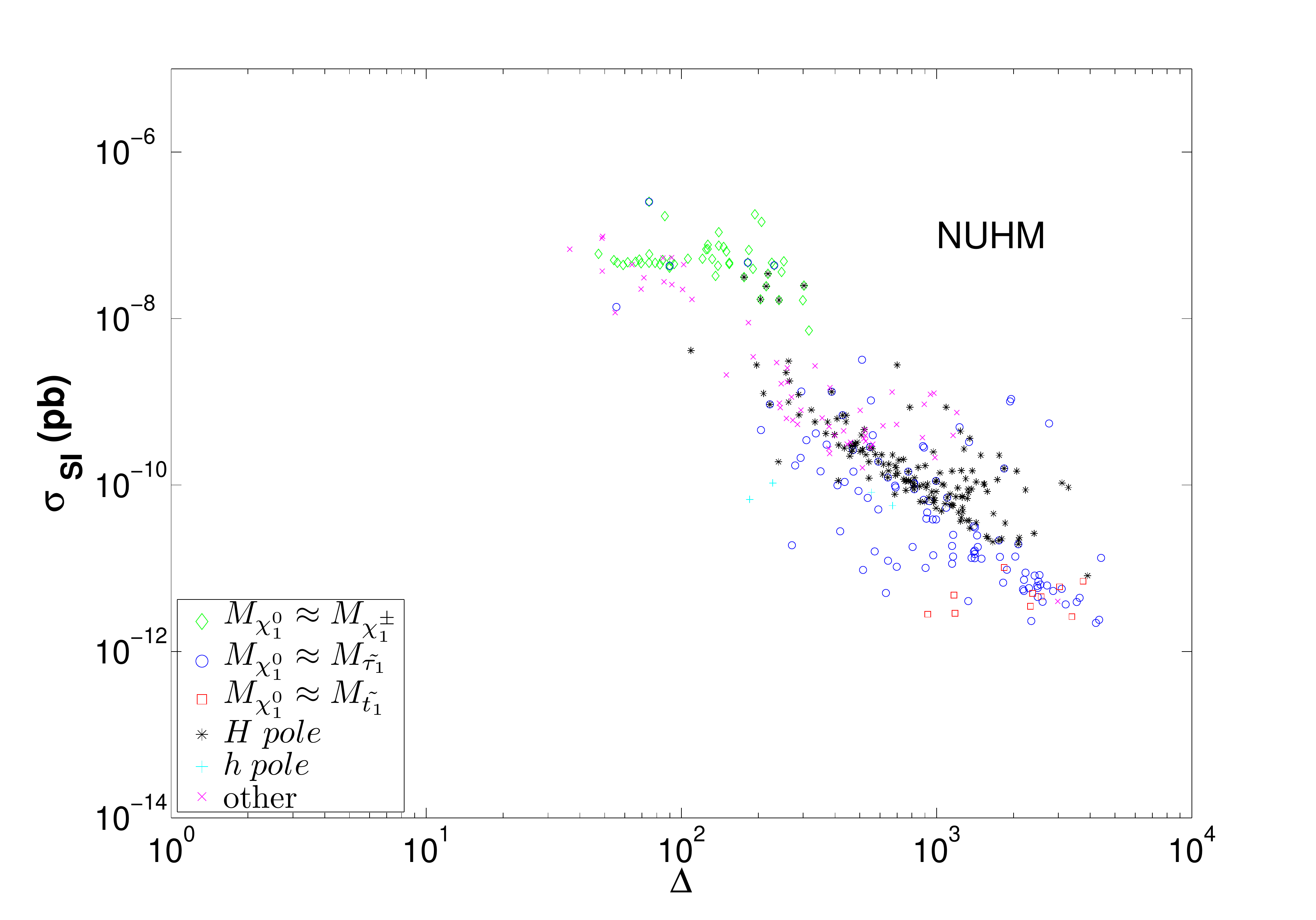}
\includegraphics[width=80mm, height=80mm]{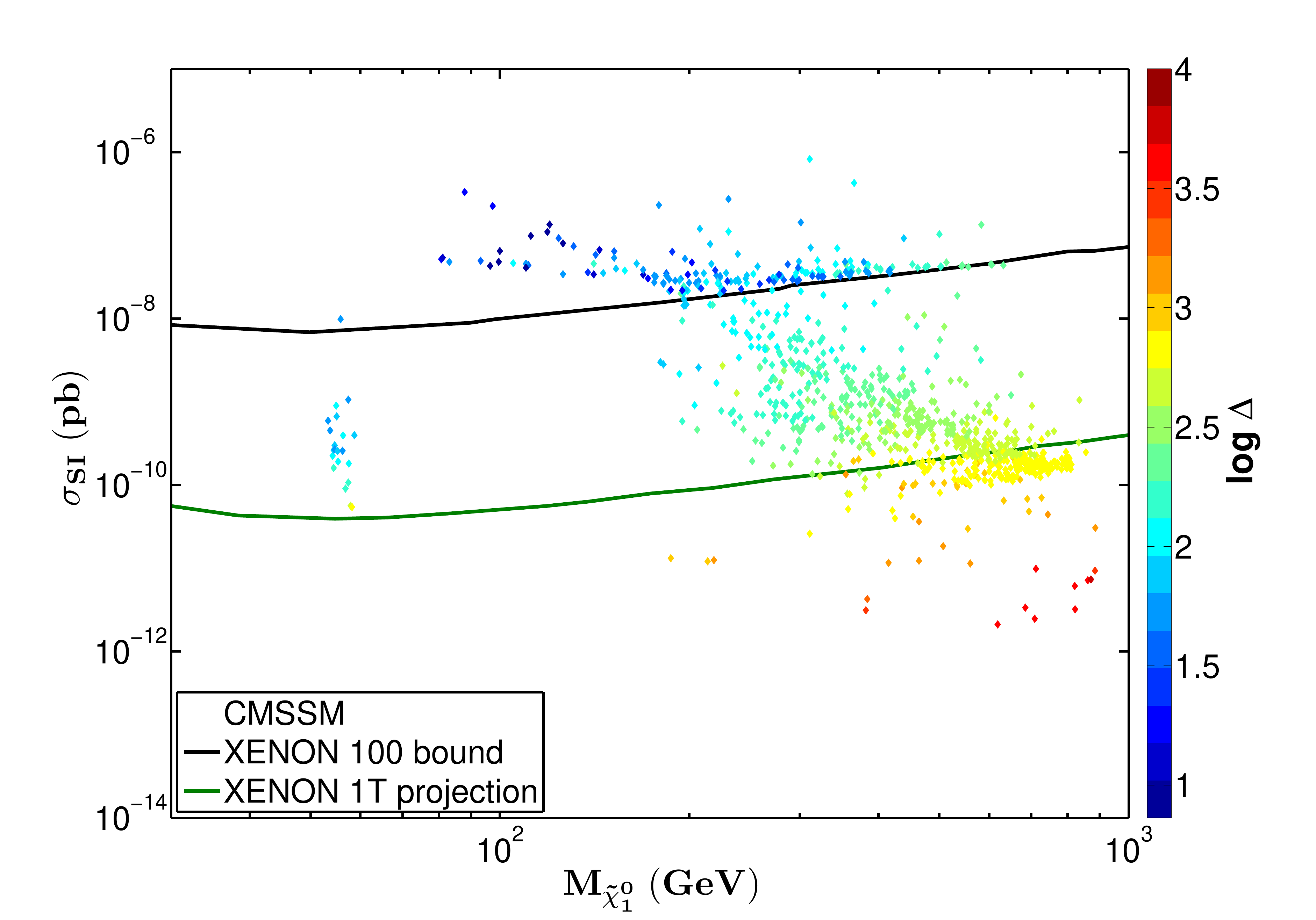}
\includegraphics[width=80mm, height=80mm]{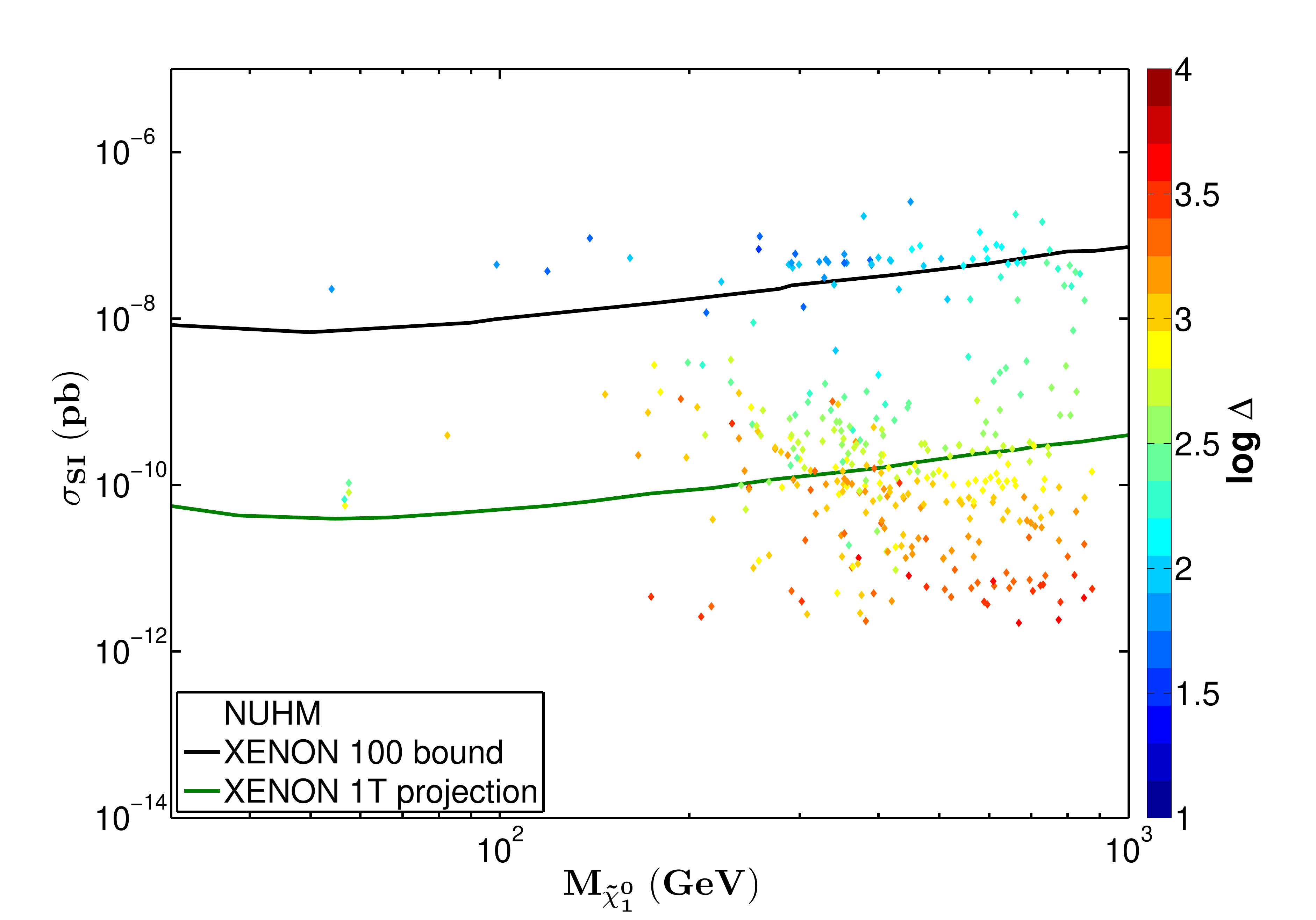}
\caption{Spin-independent cross-section for the CMSSM (left) and the NUHM (right) for neutralinos with
correct relic density $\Omega_{\na} = \Omega_{CDM}$.
In the top two panels,
 $\sigma_{SI}$ is plotted against $\Delta$; models are color-coded by mass hierarchy as shown in the
 legend. 
In the bottom two panels, $\sigma_{SI}$ is plotted against $M_{\na}$ with the Xenon-100 and
projected XENON-1T bounds as shown; models are shaded by $\Delta$.   Many of the lowest $\Delta$ points have disappeared due to the two-sided bound.}
\label{xenonbound_omega}
\end{figure}

When the two-sided bound on $\Omega_{\na}h^2$ is enforced, many points with low fine-tuning, $\Delta$, are eliminated.   From the bottom panels of Fig.~\ref{XENON_Omega}, one can see that even in the CMSSM (and moreso in the NUHM), there is significant parameter space with small $\Delta$ and $\Omega_{\na} \ll \Omega_{CDM}$. We remind the reader that these points typically have small $\sigma_{SI}$ because of the scaling necessary to compare with direct dark matter searches. When these scaled points are eliminated, a stronger correlation between $\sigma_{SI}$ and $\Delta$ emerges, even for the NUHM, as evidenced in the top panels of Fig.~\ref{xenonbound_omega}.

The implications of the results of the XENON experiment for fine-tuning are, for the most part, not qualitatively different when the second bound is enforced, as seen in the bottom panels of Fig.~\ref{xenonbound_omega}. 
For the CMSSM, the anti-correlation between fine-tuning and ease of detectability is clear.  While the general trend is still present in the NUHM, very fine-tuned points may be found at much larger $\sigma_{SI}$ and points with low fine-tuning may be found at much smaller $\sigma_{SI}$ than in the CMSSM. 
In fact, in the NUHM, low values of $\Delta \sim 200$ can have
scattering cross sections as low as $\sigma_{SI} \approx {\rm few }\times 10^{-11}$.  The lower limit
of what direct detection experiments will ever be able to probe is $\sigma_{SI} \approx 10^{-12}$ pb, below which astrophysical
neutrinos produce an irreducible background to any WIMP dark matter search~\cite{Strigari:2009bq}.

\section{Conclusion}

We have studied EWSB fine-tuning in the CMSSM and the NUHM, in light of current and upcoming direct
detection experiments.  Fine-tuning of EWSB can be approximated well as a monotonically increasing function of $\mu$.     We studied models
satisfying first a one-sided bound on the relic density $\Omega_{\na} < 0.12$ and then a two-sided bound in which
the relic density is within the $2 \sigma$ best fit of WMAP7 + BAO + H0 data.  
Our results are qualitatively similar in both cases.
We find that current direct searches for dark matter probe the least fine-tuned regions of parameter-space, or equivalently those of lowest $\mu$, and will tend to probe progressively more and more fine-tuned models, though the trend is more pronounced in the CMSSM than in the NUHM.

There is more variation in the spin-independent neutralino-nucleon elastic scattering cross section in the NUHM than in the CMSSM, especially for $m_{\na} \lesssim 150$ GeV or $m_{\na} \gtrsim 700$ GeV.  This is a consequence of the additional freedom in the Higgs sector of the NUHM.  The larger range of CP-even Higgs scalar masses in the NUHM dramatically affects the elastic scattering cross section, which is dominated by Higgs exchange.  Since the Higgs masses are bounded from below, but not bounded from above, this tends to push $\sigma_{SI}$ to lower values in the NUHM than would be expected in the CMSSM.  There is also a competing effect:  Higgsino-like dark matter is less correlated with $\Omega_{\na}$ in the NUHM than in the CMSSM, leading to significant variation in the effective scattering cross section, $\sigma_{SI}\Omega_{\na}/\Omega_{CDM}$.  Unless the LSP is purely higgsino (a case which occurs only in the NUHM and not in the CMSSM), this effect tends to push $\sigma_{SI}$ to larger values.
In general, the result is an expanded range of viable neutralino-nucleon scattering cross sections in the NUHM relative to that in the CMSSM, and a lower level of correlation between the degree of fine-tuning and direct detection prospects.

Additionally, we examined the relationship between electroweak fine-tuning and SUSY mass hierarchy, studying the specific cases of $M_{\na} \approx M_{\cha}, M_{\na} \approx M_{\sta}, M_{\na} \approx M_{\ta}$, the light and heavy Higgs
poles, and any additional models classified as ``other''.  
Requiring only that neutralino dark matter make up some fraction of the dark matter in the Universe, we find that XENON-100 has already ruled out a significant fraction of parameter space in the CMSSM with low fine-tuning, but a less significant chunk of the NUHM.  In both cases, models with $m_{\na}\approx m_{\cha}$ may have low fine-tuning but large gaugino masses.
For $M_{\na} \sim M_{\sta}$, in the CMSSM
most cases with very light $m_{\na} \approx m_{\sta} \lesssim 200$ GeV would be accessible to XENON-1T, and all cases are well above the irreducible neutrino background at $\sigma_{SI} \approx 10^{-12}$ pb.
However, for the NUHM,
it is possible that the lightest neutralino has $\sigma_{SI} \lesssim 10^{-12}$ pb for a large range of $m_{\na}$. 
For the case of $M_{\na} \sim M_{\ta}$, 
a low $\ta$ mass is easily detectable at the LHC, but it is clear that the neutralino dark matter would not be
discoverable even with XENON-1T.   Furthermore one can see that almost all of the points in this case are 
quite fine-tuned with $\Delta > 1000$. 

When we apply the two-sided bound on the relic density, some of the least fine-tuned (lowest $\Delta$) points
do not survive.  The implications of the results of the Xenon experiment for fine-tuning are, for the most part, not qualitatively different when the lower bound is enforced.
In the CMSSM, if neutralino LSPs are light or have small $\Delta < 200$, then
they will be seen or ruled out by the next generation direct detection scattering experiments such as XENON-1T.
For the NUHM, however, models with low values of $\Delta \sim 200$ may evade detection by XENON-1T.

\vspace{3cm}
\noindent {\it Acknowledgements:} 
 This research is  supported in
part by Department of Energy (DOE) grant  DE-FG02-95ER40899 and  by the Michigan Center
for Theoretical Physics.  
P.S. is supported by the National Science Foundation under Grant Number PHY-0969020 and by the University of Utah.
KF thanks the Texas Cosmology Center (TCC) where she is a Distinguished Visiting Professor. TCC is supported by the College of Natural Sciences and the Department of Astronomy at the University of Texas at Austin and the McDonald Observatory. KF  also thanks the  Aspen Center for Physics for hospitality during her visit.  Additionally, we thank Daniel Feldman for helpful comments.\\

\end{document}